\documentclass[10pt,journal,compsoc]{IEEEtran}

\usepackage{bibnames}
\usepackage{amssymb}
\usepackage{verbatim}
\usepackage{amsmath}
\usepackage{amsthm}
\usepackage{graphicx}
\usepackage[font=scriptsize]{caption}
\usepackage{subcaption}
\usepackage[linesnumbered, ruled,vlined]{algorithm2e}
\usepackage{listings}
\usepackage[flushleft]{threeparttable}
\usepackage{multirow}
\usepackage{listings}
\usepackage{cite}
\usepackage{hyperref}
\usepackage{mdframed}
\usepackage{tabularx}
\usepackage{flushend}
\usepackage[T1]{fontenc}
\usepackage[switch]{lineno}
\usepackage{enumitem}

\usepackage{balance}

\usepackage[table]{xcolor}
\definecolor{ao}{rgb}{0.0, 0.5, 0.0}
\newtheorem{definition}{Definition}[section]
\newtheorem{property}{Property}[section]

\newcommand{\tool}{\textsc{VarCop}\xspace}

\newcommand{\spc}{\textit{Suspicious PC}\xspace}
\newcommand{\spcs}{\textit{Suspicious PCs}\xspace}
\newcommand{\bpc}{\textit{Buggy PC}\xspace}
\newcommand{\bpcs}{\textit{Buggy PCs}\xspace}

\begin{document}

\title{A Variability Fault Localization Approach for Software Product Lines}

\author{Thu-Trang~Nguyen,
        Kien-Tuan~Ngo,
        Son~Nguyen,
        and~Hieu~Dinh~Vo % stops a space
\IEEEcompsocitemizethanks{
    \IEEEcompsocthanksitem Thu-Trang~Nguyen, Kien-Tuan~Ngo, Son~Nguyen, and Hieu~Dinh~Vo are with Department of Software Engineering, Faculty of Information Technology, University of Engineering and Technology, Vietnam National University, Hanoi, Vietnam.
    \IEEEcompsocthanksitem Corresponding author: Hieu~Dinh~Vo. \protect\\
    E-mail: \href{mailto:hieuvd@vnu.edu.vn}{hieuvd@vnu.edu.vn}.
    }
}

\IEEEtitleabstractindextext{%

\begin{abstract}
Software fault localization is one of the most expensive, tedious, and time-consuming activities in program debugging. This activity becomes even much more challenging in Software Product Line (SPL) systems due to variability of failures.
These unexpected behaviors are induced by variability faults which can only be exposed under some combinations of system features. The interaction among these features causes the failures of the system. 
Although localizing bugs in single-system engineering has been studied in-depth, variability fault localization in SPL systems still remains mostly unexplored.
In this article, we present \tool, a novel and effective variability fault localization approach. 
For an SPL system failed by variability bugs, \tool isolates suspicious code statements by analyzing the overall test results of the sampled products and their source code. The isolated suspicious statements are the statements related to the interaction among the features which are necessary for the visibility of the bugs in the system.
In \tool, the suspiciousness of each isolated statement is assessed based on both the overall test results of the products containing the statement as well as the detailed results of the test cases executed by the statement in these products. 
On a large public dataset of buggy SPL systems, our empirical evaluation shows that \tool significantly improves two state-of-the-art techniques by 33\% and 50\% in ranking the incorrect statements in the systems containing a single bug each.
In about two-thirds of the cases, \tool correctly ranks the buggy statements at the top-3 positions in the ranked lists. 
For the cases containing multiple bugs, 
\tool outperforms the state-of-the-art approaches 2 times and 10 times in the proportion of bugs localized at the top-1 positions. Especially, in 22\% and 65\% of the buggy versions, \tool correctly ranks at least one bug in a system at the top-1 and top-5 positions. 
\end{abstract}

\begin{IEEEkeywords}
Fault localization, variability bugs, feature interaction, software product line, configurable code
\end{IEEEkeywords}}

\maketitle

\IEEEdisplaynontitleabstractindextext

\IEEEpeerreviewmaketitle

\IEEEraisesectionheading{\section{Introduction}}

% Software Product Line
\IEEEPARstart{M}any software projects enable developers to configure to different environments and requirements. In practice, a project adopting the Software Product Line (SPL) methodology~\cite{SPLBook} can tailor its functional and nonfunctional properties to the
% needs and 
requirements of users~\cite{SPLBook, apel2016feature}. This has been done using a very large number of \textit{options} which are used to control different  \textit{features}~\cite{apel2016feature} additional to the \textit{core software}. A set of \textit{selections} of all the features (\textit{configurations}) defines a program \textit{variant} (\textit{product}). For example, Linux Kernel supports thousands of features controlled by +12K compile-time options, that can be configured to generate specific kernel \textit{variants} for billions of possible scenarios.

%
%QA in SPL & Variability Bug and Variability FL (VFL)
However, the variability that is inherent to SPL systems challenges quality assurance (QA)~\cite{garvin2011feature, apel2013strategies, 42bugs, sampling_comparision, ase19prioritization}. 
In comparison with the single-system engineering, fault detection and localization through testing in SPL systems are more problematic, as a bug can be \textit{variable} (so-called \textit{variability bug}), which can only be exposed under \textit{some} combinations of the system features~\cite{garvin2011feature, interaction_complexity}. Specially, there exists a set of the features that must be selected to be on and off together to necessarily reveal the bug. 
%
% We call this set of the feature selections a \textit{Buggy \textbf{P}artial \textbf{C}onfiguration (\bpc)}.
%
%
Due to the presence/absence of the \textit{interaction} among the features in such set,
% in the \bpc, 
the buggy statements behave differently in the products where these features are on and off together or not.
%the \bpc. 
Hence,\textit{ the incorrect statements can only expose their bugginess in some particular products, yet cannot in the others.}
Specially, in an SPL system, variability bugs only cause failures in certain products, and the others still pass all their tests.
% Thus, the variability bug only causes the failures in the products with a \bpc. 
%
This variability property causes considerable difficulties for localizing this kind of bugs in SPL systems.
In the rest of this paper, variability bugs are our focus, and we simply call a system containing variability bugs a \textit{buggy (SPL) system}.
% especially in the large SPL systems such as Linux kernel or Mozilla Firefox.

% Current approaches and their problems 
Despite the importance of \textit{variability fault localization}, the existing fault localization (FL) approaches~\cite{wong2016survey, arrieta2018spectrum, pearson2017evaluating} are not designed for this kind of bugs. These techniques are specialized for finding bugs in a particular product. 
For instance, to isolate the bugs causing failures in multiple products of a single SPL system, the slice-based methods~\cite{static_slicing, dynamic_slicing, wong2016survey} could be used to identify all the failure-related slices for each product independently of others. 
Consequently, there are multiple sets of large numbers of isolated statements that need to be examined to find the bugs.  
%
% %
This makes the slice-based methods~\cite{wong2016survey} become impractical in SPL systems.

In addition, the state-of-the-art technique, Spectrum-Based Fault Localization (SBFL)~\cite{pearson2017evaluating, keller2017critical, naish2011model, abreu2009spectrum, abreu2007accuracy} can be used to calculate the suspiciousness scores of code statements based on the test information (i.e., program spectra) of each product of the system separately. For each product, it produces a ranked list of suspicious statements. As a result, there might be multiple ranked lists produced for a single SPL system which is failed by variability bugs. 
From these multiple lists, developers cannot determine a starting point to diagnose the root causes of the failures. Hence, it is inefficient to find variability bugs by using SBFL to rank suspicious statements in multiple variants separately.

%Hence, SBFL cannot be directly applied for variability bugs.

%Adaptation
Another method to apply SBFL for localizing variability bugs in an SPL system is that one can treat the whole system as a single program~\cite{ourdataset}. This means that the mechanism controlling the presence/absence of the features in the system (e.g., the preprocessor directives \texttt{\#ifdef}) would be considered as the corresponding conditional \texttt{if-then} statements during the localization process.
% and considering the states of the features as test inputs.
%
By this adaptation of SBFL, a single ranked list of the statements for variability bugs can be produced according to the suspiciousness score of each statement. 
%
%
% Specially, in product-based testing, a set of products in an SPL system are sampled and tested individually. Each sampled product has its own test suite. Even for a test case is designed to test functionality of a feature in domain engineering, when products are instantiated in application engineering, the test is also made as concrete test cases according to the particular requirement of each product~\cite{do2012strategies}. 
%
Note that, we consider the product-based testing~\cite{do2012strategies, thum2014classification}. Specially, each product is considered to be tested individually with its own test set. Additionally, a test, which is  designed to test a feature in domain engineering, is concretized to multiple test cases according to products' requirements in application engineering~\cite{do2012strategies}. 
Using this adaptation, the suspiciousness score of the statement is measured based on the total numbers of the passed and failed tests executed by it in all the tested products. 
% Moreover, Arrieta et al.~\cite{arrieta2018spectrum} also propose a method for localizing bugs in SPL systems at the feature-level by adapting SBFL.They consider each product as a test (i.e., passed tests are passing products and failed tests are failing products), and the program spectra record the selections of the features in the products. By this approach, each feature is considered as a fault localization unit, SBFL is used to calculate suspiciousness for each feature, and all the statements in the same feature will have the same suspiciousness scores.
%
Meanwhile, the characteristics including the interactions between system features and the variability of failures among products are also useful to isolate and localize variability bugs in SPL systems.
However, these kinds of important information are not utilized in the existing approaches. 

In this paper, we propose {\tool}, a novel fault localization approach for variability bugs. Our key ideas in {\tool} is that variability bugs are localized based on (i) the interaction among the features which are necessary to reveal the bugs, and (ii) the bugginess exposure which is reflected via both the overall test results of products and the detailed result of each test case in the products.
%

% In this paper, we propose {\tool}, a novel fault localization approach for variability bugs. Our key ideas in {\tool} is that variability bugs are localized by (1) identifying the cause of the bugs' visibility which is the interaction among the features which are necessary to be on or off to reveal the bugs, and (2) utilizing both the overall test results of products and the detailed result of each test in products which reflect how the bugginess expose.

Particularly, for a buggy SPL system, {\tool} detects sets of the features which need to be selected on/off together to make the system fail by analyzing the overall test results (i.e., the state of passing all tests or failing at least one test) of the products. 
We call each of these sets of the feature selections a \textit{Buggy \textbf{P}artial \textbf{C}onfiguration (\bpc)}. Then, \tool analyzes the interaction among the features in these \bpcs to isolate the statements which are suspicious. 
In \tool, the suspiciousness of each isolated statement is assessed based on two criteria.
The first criterion is based on the overall test results of the products containing the statement. By this criterion, the more failing products and the fewer passing products  where the statement appears, the more suspicious the statement is.
%
% Meanwhile, the second one is assessed based on the detailed results of every test case executed by the statement in the failing products containing that statement. 
Meanwhile, the second one is assessed based on the suspiciousness of the statement in the failing products which contain it. Specially, in each failing product, the statement's suspiciousness is measured based on the detailed results of the products' test cases.
The idea is that if the statement is more suspicious in the failing products based on their detailed test results, the statement is also more likely to be buggy in the whole system.
We conducted experiments to evaluate {\tool} in both single-bug and multiple-bug settings on a dataset of 1,570 versions (cases) containing variability bug(s)~\cite{ourdataset}. We compared {\tool} with the state-of-the-art approaches including (SBFL)~\cite{pearson2017evaluating, keller2017critical, naish2011model, abreu2009spectrum, abreu2007accuracy}, the combination of the slicing method and SBFL (S-SBFL)~\cite{chaleshtari2020smbfl, li2020more}, and Arrieta et al.~\cite{arrieta2018spectrum} using 30 most popular SBFL ranking metrics~\cite{keller2017critical, naish2011model, pearson2017evaluating}. 

For the cases containing a single incorrect statement (single-bug), our results show that {\tool} significantly outperformed S-SBFL, SBFL, and Arrieta et
al.~\cite{arrieta2018spectrum} in \textbf{all 30/30} metrics by \textbf{33\%}, \textbf{50\%}, and \textbf{95\%} in \textit{Rank}, respectively. Impressively, {\tool} correctly ranked the bugs at the top-3 positions in \textbf{+65\%} of the cases. 
% Meanwhile, S-SBFL, SBFL, and Arrieta et al.~\cite{arrieta2018spectrum} ranked them at top-3 positions in only 50\% and 1\% of the cases, respectively. 
In addition, \tool effectively ranked the buggy statements first in about \textbf{30\%} of the cases, which \textbf{doubles} the corresponding figure of SBFL.

For localizing multiple incorrect statements (multiple-bug), after inspecting the first statement in the ranked list resulted by \tool, up to \textbf{10\%} of the bugs in a system can be found, which is \textbf{2 times} and \textbf{10 times} better than S-SBFL and SBFL, respectively.
% Moreover, \tool correctly ranked \textbf{one-third} of the bugs in a case at the top-5 positions. 
Especially, our results also show that in \textbf{22}\% and \textbf{65\%} of the cases, \tool effectively localized at least one buggy statement of a system at top-1 and top-5 positions. From that, developers can iterate the process of bugs detecting, bugs fixing, and regression testing to quickly fix all the bugs and assure the quality of SPL systems.

In brief, this paper makes the following contributions:

\begin{enumerate}
    \item A formulation of \textit{Buggy Partial Configuration (\bpc)} where the interaction among the features in the \bpc is the root cause of the failures caused by variability bugs in SPL systems.
    \item {\tool}: A novel effective approach/tool to localize variability bugs in SPL systems.
    \item An extensive experimental evaluation showing the performance of {\tool} over the state-of-the-art methods.
\end{enumerate}
 
\section{Motivating Example}
\label{sec:example_approach}
In this section, we illustrate the challenges of localizing variability bugs and motivate our solution via an example.

\subsection{An Example of Variability Bugs in SPL Systems}
\begin{figure}
    \centering
\begin{lstlisting}[language=Java]
int maxWeight = 2000, weight = 0;
int maxPersons = 20; 
//#ifdef Empty
void empty(){ persons.clear();}
//#endif
void enter(Person p){
    persons.add(p);
    //#ifdef Weight
    weight += p.getWeight();
    //#endif
}
void leave(Person p){
    persons.remove(p);
    //#ifdef Weight
    weight -= p.getWeight();
    //#endif		
}                                                                                                                                               
ElevState stopAtAFloor(int floorID){	
    ElevState state = Elev.openDoors;
    boolean block = false;
    for (Person p: new ArrayList<Person>(persons)) 
        if (p.getDestination() == floorID)  
            leave(p);	
    for (Person p : waiting)    enter(p);
    //#ifdef TwoThirdsFull
    if ((weight == 0                                             && persons.size() >= maxPersons*2/3)                 || weight >= maxWeight*2/3)
        block = true;
    //#endif
    //#ifdef Overloaded
    if(block == false){
        if ((weight == 0                                       && persons.size() >= maxPersons)                 || weight == maxWeight)
            //Patch: weight >= maxWeight
            block = true;
    }
    //#endif
    if (block == true)
        return Elev.blockDoors;
    return Elev.closeDoors;
}
\end{lstlisting}
   \caption{An example of variability bug in Elevator System}
    \label{fig:example_code}
\end{figure}{}

Fig. \ref{fig:example_code} shows a simplified variability bug in \textit{Elevator System}~\cite{ourdataset}. The overall test results of the sampled products are shown in Table \ref{Table:example_test_result}. In Fig. \ref{fig:example_code}, the bug (incorrect statement) at line 31 causes the failures in products $p_6$ and $p_7$. 

%Specification of the example (What the example supposed to do)
\textit{Elevator System} is expected to simulate an elevator and consists of 5 features:
\textit{Base, Empty, Weight, TwoThirdsFull,} and \textit{Overloaded}. Specially, \textit{Base} is the mandatory feature implementing the basic functionalities of the system, while the others are optional. 
%
% \textit{Empty}
% (line 38) 
% resets the state of the elevator. 
%
% For \textit{Weight}, when it is on,
% (lines 6 and 12) 
% the elevator's load will be tracked as its total loaded weight, otherwise the load is calculated based on the number of the people inside the elevator cabin. 
%
\textit{TwoThirdsFull} is expected to
% (line 23) 
limit the load not to exceed 2/3 of the elevator's capacity, while \textit{Overloaded} 
% (line 27) 
ensures the maximum load is the elevator's capacity.

%About the actual implementation
However, the implementation of \textit{Overloaded} (lines 30--34) does not behave as specified. 
If the total loaded weight (\texttt{weight}) of the elevator is tracked, then instead of blocking the elevator when \texttt{weight} exceeds its capacity (\texttt{weight >= maxWeight}), its actual implementation blocks the elevator \textbf{only} when \texttt{weight} is equal to \texttt{maxWeight} (line 31). Consequently, if \textit{Weight} and \textit{Overloaded} are on (and \textit{TwoThirdsFull} is off), even the total loaded weight is greater than the elevator's capacity, then (\texttt{block==false}) the elevator still dangerously works without blocking the doors (lines 37--39). 

\begin{table}
\scriptsize
\centering
\caption{The sampled products and their overall test results}
\label{Table:example_test_result}
\begin{threeparttable}
\begin{tabular}{|c|c|c|c|c|c|c|}
\hline
$P$ & $C$ & \textit{Base} & \textit{Empty} & \textit{Weight}   & \textit{TwoThirdsFull} & \textit{Overloaded}  \\\hline
$p_1$   &   $c_1$    & T     & F     & T            & F             & F           \\\hline
$p_2$   &   $c_2$    & T    & T      & T                & F             & F           \\\hline
$p_3$   &  $c_3$     & T      & T     & F            & F             & F           \\\hline
$p_4$   &  $c_4$    & T      & F     &T           & T             & F             \\\hline
$p_5$   &  $c_5$    & T      & F     & T           & T             & T           \\\hline
\rowcolor[HTML]{C0C0C0} 
$p_6$   & $c_6$   & T        &T      & T                & F             & T           \\\hline
\rowcolor[HTML]{C0C0C0} 
$p_7$ &  $c_7$     &T       & F     &T           & F             & T           \\\hline
\end{tabular}
\begin{tablenotes}
    \item $P$ and $C$ are the sampled sets of products and configurations. 
    \item $p_6$ and $p_7$ fail at least one test (\textit{failing products}). Other products pass all their tests (\textit{passing products}).
\end{tablenotes}
\end{threeparttable}
\end{table}

%About the bug and fixing this bug
This bug (line 31) is \textit{variable} (\textit{variability bug}). It is revealed not in all the sampled products, but only in $p_6$ and $p_7$ (Table \ref{Table:example_test_result}) due to the \textit{interaction} among \textit{Weight}, \textit{Overloaded}, and \textit{TwoThirdsFull}. 
Specially, the behavior of \textit{Overloaded} which sets the value of \texttt{block} at line 33 is interfered by \textit{TwoThirdsFull} when both of them are on (lines 27 and 30). Moreover, the incorrect condition at line 31 can be exposed only when \textit{Weight = T}, \textit{TwoThirdsFull=F}, and \textit{Overloaded = T} in $p_6$ and $p_7$. 
Hence, understanding the root cause of the failures to localize the variability bug could be very challenging.

\subsection{Observations}
\label{subsec:observations}
% % 

%OBSERVATION 1
% \textbf{O1.} 
For an SPL system containing variability bugs, there are certain features that are (ir)relevant to the failures~\cite{98bugs,mordahl2019empirical,garvin2011feature}. 
In Fig. \ref{fig:example_code}, enabling or disabling feature \textit{Empty} does not affect the failures. Indeed, some products still fail ($p_6$ and $p_7$) or pass their test cases ($p_1$ and $p_2$) regardless of whether \textit{Empty} is on or off. 
Meanwhile, there are several features which \textbf{must be enabled/disabled together} to make the bugs visible. In other words, for certain products, changing their current configurations by switching the current selection of anyone in these relevant features makes the resulting products' overall test results change. 
For \textit{TwoThirdsFull}, switching its current off-selection in the failing product, $p_7$, makes the resulting product, $p_5$, where $TwoThirdsFull=T$, behave as expected (Table~\ref{Table:example_test_result}). 
The reason is that in $p_5$, the presence of \textit{Weight} and \textit{TwoThirdsFull} impacts \textit{Overloaded}, and consequently \textit{Overloaded} does not expose its incorrectness. 
%confirm observation 1
%
% In addition, this has been confirmed by the Variability Bugs Database (VBDb)~\cite{98bugs} - the collection of 98 real world variability bugs. In VBDb, there are 41 bugs revealed by a single feature and the remaining 57 bugs involved 2--5 features. Moreover, the occurrence condition of these bugs is relevant to  both enabled and disabled features. For instance, 49 bugs occurred when all of the relevant features are enabled, meanwhile, the presence condition of the other half of the bugs is that at least one relevant feature is disabled. For each variability bug, the impact of the relevant features on each other makes the bug exposed or concealed.
%
In fact, this characteristic of variability bugs has been confirmed by the public datasets of real-world variability bugs~\cite{98bugs,mordahl2019empirical}. For example, in the VBDb~\cite{98bugs}, there are 41/98 bugs revealed by a single feature and the remaining 57/98 bugs involved 2--5 features. The occurrence condition of these bugs is relevant to both enabled and disabled features. Particularly, 49 bugs occurred when all of the relevant features must be enabled. Meanwhile, the other half of the bugs require that at least one relevant feature is disabled. 
% For each variability bug, the impact of the relevant features on each other makes the bug exposed or concealed.

%
In fact, the impact of features on each other is their \textbf{feature interaction}~\cite{apel2013exploring, garvin2011feature}. The presence of a feature interaction  makes certain statements behave differently from when the interaction is absent.
For variability bugs, the presence of the interaction among the relevant features\footnote{Without loss of generality, for the cases where there is only one relevant feature, the feature can impact and interact itself.} exposes/conceals the bugginess of the statements that cause the unexpected/expected behaviors of the failing/passing products~\cite{98bugs,garvin2011feature}.
Thus, to localize variability bugs, it is necessary to identify such sets of the relevant features as well as the interaction among the features in each set, which is the root cause of the failures in failing products. 

In an SPL system containing variability bugs, there might be multiple such sets of the relevant features. Let consider a particular set of relevant features.
%In an SPL system containing variability bugs, there might be multiple sets of the relevant features. Let consider a particular set of relevant features whose selections are appropriately enabled and disabled together.

%OBSERVATION 2
\textbf{O1.} 
\textit{In the features $F_{E}$ which \textbf{must be enabled together} to make the bugs visible, the statements that implement the interaction among these features provide valuable suggestions to localize the bugs.}
For instance, in Fig. \ref{fig:example_code}, $F_{E}$ consists of \textit{Weight} and \textit{Overloaded}, and the interaction between these features contains the buggy statement at line 31 ($s_{31}$) in \textit{Overloaded}. This statement uses variable \texttt{weight} defined/updated by feature \textit{Weight} (lines 9 and 15).
Hence, detecting the statements that implement the interaction among the features in $F_{E}$ could provide us valuable indications to localize the variability bugs in the systems.

%OBSERVATION 3
\textbf{O2.} 
% Moreover, \textit{in the features $F_D$ which \textbf{must be disabled together} to reveal the bugs, when they are enabled (if possible), the statements, which impact the interaction among the features in $F_{E}$, also provide useful indications to help us find bugs}. 
Moreover, \textit{in the features $F_D$ which \textbf{must be disabled together} to reveal the bugs, the statements impacting the interaction among the features in $F_{E}$ (if the features in $F_D$ and $F_E$ are all on), also provide useful indications to help us find bugs}.
In Fig. \ref{fig:example_code}, although the statements from lines 26--27 in \textit{TwoThirdsFull} (being disabled) are not buggy, analyzing the impact of these statements on the interaction between \textit{Overloaded} and \textit{Weight} can provide the suggestion to identify the buggy statement. 
The intuition is that the features in $F_{D}$ have the impacts of ``hiding" the bugs when these features are enabled. 
In this example, when \textit{Weight}, \textit{TwoThirdsFull} and \textit{Overloaded} are all on, if the loaded weight exceeds \texttt{maxWeight*2/3}  (i.e., the conditions at line 26 are satisfied), then \texttt{block = true}, and the statements from 31--33 (in \textit{Overloaded}) cannot be executed. As a result, the impact of the incorrect condition at line 31 is ``hidden".
Thus, we should consider the impact of the features in $F_{D}$ (as if they are on) on the interaction among the features in $F_{E}$ in localizing variability bugs.

%About scoring
\textbf{O3.} 
%Local + global measurement
%
% The tests of a particular product are specialized to verify its behaviors.
% %
% For example, since the actual and expected behaviors of $p_5$ are different from $p_6$'s, the test cases of $p_5$ are incomparable to the test cases of $p_6$. As a result, the tests are used to verify $s_{31}$ in $p_5$, where the statements in lines 26--27 ($s_{26}$--$s_{27}$) are present, should not be used to verify $s_{31}$ in $p_6$, in which  $s_{26}$--$s_{27}$ are absent.
% Hence, to accurately assess the suspiciousness of statements, \textit{the program spectra of a particular product should be used to measure the suspiciousness of the statements in that product only.}
% %
% %
% Moreover, in Fig. \ref{fig:example_code}, the products containing $s_{31}$ are not only failing products ($p_6$ and $p_7$), but also a passing one, $p_5$.
% %
% Furthermore, the bugginess of $s_{31}$ is exposed differently in via the detailed (case-by-case) test results in the failing products, $p_6$ and $p_7$.
% %
% Thus, to holistically assess the suspiciousness of a statement $s$, \textit{the score of $s$ should reflect the statement's suspiciousness in the overall test result of the products which contain $s$, as well as the suspiciousness of $s$ via the detailed test results in all failing products where $s$ contributes to.} 
%
% %
%
For a buggy SPL system, a statement can appear in both failing and passing products. Meanwhile, the states of failing or passing of the products expose the bugginess of the contained buggy statements. Thus, \textit{the overall test results of the sampled products can be used to measure the suspiciousness of the statements.}
Furthermore, the bugginess of an incorrect statement can be exposed via the detailed (case-by-case) test results in every failing product containing the bug.
In our example, $s_{31}$ is contained in two failing products $p_6$ and $p_7$, and its bugginess is expressed differently via the detailed test results in $p_6$ and $p_7$.
Thus, to holistically assess the suspiciousness of a statement $s$, \textit{the score of $s$ should also reflect the statement's suspiciousness via the detailed test results in every failing product where $s$ contributes to.} 

Among these observations, \textbf{O1} and \textbf{O2} will be theoretically discussed in Section \ref{sec:root_cause}. We also empirically validated observation \textbf{O3} in our experimental results (Section \ref{sec:empirical_results}).

\begin{figure}
    \centering
    \includegraphics[width=1\linewidth]{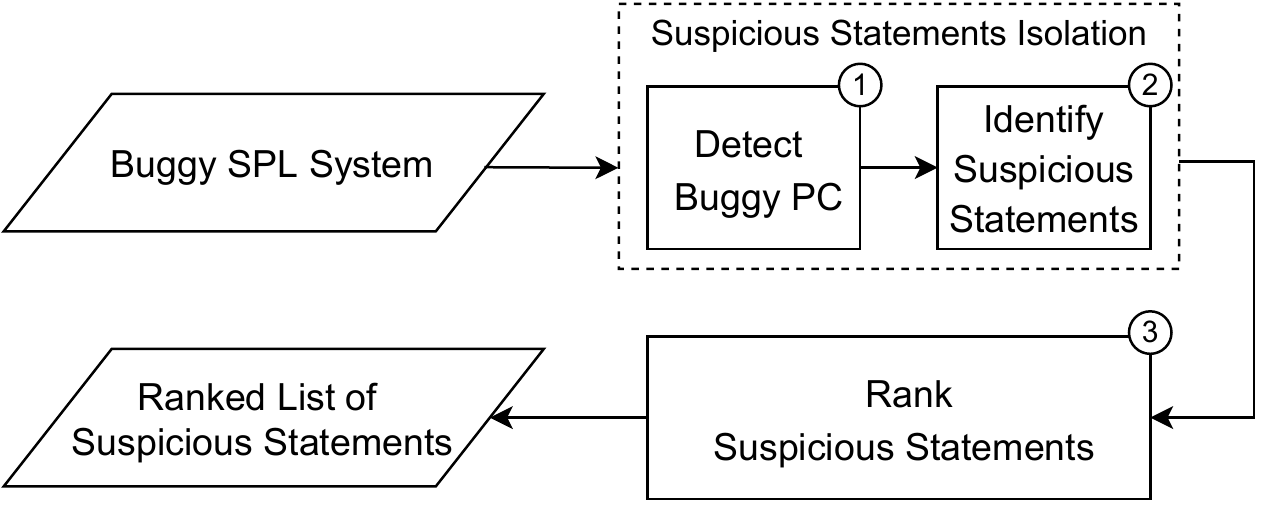}
    \caption{\tool's Overview}
    \label{fig:varcop_overview}
\end{figure}

\subsection{{\tool} Overview}
Based on these observations, we propose {\tool}, a novel  variability fault localization approach. %
For a given buggy SPL system, the input of {\tool} consists of a set of the tested products and their program spectra. 
\tool outputs a ranked list of suspicious statements in three steps (Fig. \ref{fig:varcop_overview}):
\begin{enumerate}
    \item First, by analyzing the configurations and the overall test results of the sampled products, {\tool} detects minimal sets of features whose \textit{selections} (the states of being \textit{on/off}) make the bugs (in)visible. Let us call such a set of selections a \textit{Buggy Partial Configuration (\bpc)}. In Fig. \ref{fig:example_code}, $\{Weight=T$, $Overloaded=T$, $TwoThirdsFull=F\}$ is a \bpc. 
    
    \item Next, for each failing product, {\tool} aims to isolate the \textit{suspicious statements} which are responsible for implementing the interaction among the features in each detected \bpc.
    Specially, the feature interaction implementation is a set of the statements which these features use to \textit{impact} each other.
    % which potentially cause the failures. 
    For example, in $p_7$, {\tool} analyzes its code to detect the implementation of the interaction among \textit{Weight}, \textit{Overloaded}, and \textit{TwoThirdsFull} (\textbf{O1} and \textbf{O2}),
    %
    % In $p_7$ the statements implementing the interaction among these three features 
    and this interaction implementation includes the statements at lines 9, 15, and 31. Intuitively, all the statements in $p_7$ which have an impact on these statements or are impacted by them are also suspicious.
    
    % \item Finally, the suspicious statements are ranked by considering both their suspiciousness levels in the failing products and their contributions in the passing products. Particularly, for each detected suspicious statement, in every failing product, the statement is assigned a \textit{local suspiciousness score} which is computed by considering the program spectra of the product. Next, the local scores of the statement in the failing products are normalized and systematically aggregated to form its \textit{global suspiciousness score} in the whole system (\textbf{O3}).
    
    \item Finally, the suspicious statements are ranked by examining how their suspiciousness exposes in both the overall test results of the containing products (\textit{product-based suspiciousness assessment}) and these products' detailed case-by-case test results (\textit{test case-based suspiciousness assessment}). Particularly, for each isolated statement, the product-based assessment is calculated based on the numbers of the passing and failing products containing the statement.
    Meanwhile, the test case-based suspiciousness is assessed by aggregating the suspiciousness scores of the statement in the failing products which are calculated based on the detailed results of the tests executed by the statement.  
    %
    % Meanwhile, the test case-based assessment value is aggregated from the suspiciousness assessments of all the failing products which are calculated independently based on the detailed results of the tests executed by the statement.
    (\textbf{O3}).

\end{enumerate}

\section{Concepts and Problem Statement}
\label{Sec:concept}

A software product line is a product family that consists of a set of products sharing a common code base. These products distinguish from the others in terms of their \textit{features}~\cite{SPLBook}. 
%
% A feature is an optional or incremental unit of functionalities. 
% In this paper, the concept of SPL is defined as follows:

\begin{definition}{\textbf{(Software Product Line System).}}
\label{Def:system}
A \textbf{S}oftware \textbf{P}roduct \textbf{L}ine System (SPL) $\mathfrak{S}$ is a 3-tuple $\mathfrak{S}=\langle\mathbb{S}, \mathbb{F}, \varphi \rangle$, where:
\begin{itemize}
    \item $\mathbb{S}$ is a set of code statements that are used to implement $\mathfrak{S}$.
    \item $\mathbb{F}$ is a set of the features in the system. A \textit{feature selection} of a feature $f \in \mathbb{F}$ is the state of being either enabled (on) or disabled (off) ($f = T/F$ for short).
    \item $\varphi: \mathbb{F} \to 2^\mathbb{S}$ is the feature implementation function. For a feature $f\in \mathbb{F}$, $\varphi(f) \subset \mathbb{S}$ refers to the implementation of $f$ in $\mathfrak{S}$, and $\varphi(f)$ is included in the products where $f$ is on.
\end{itemize}
\end{definition} 

% In our example in Fig. \ref{fig:example_code}, the set of features, $\mathbb{F}$, of the Elevator system is $\mathbb{F}$ = $\{Base, Empty, TwoThirdsFull,$ $Weight, Overloaded\}$, and each feature has its own implementation. 
%
% For example, $\varphi(Overloaded)$ = $\{s_{30}, s_{31}, s_{33}\}$.
% or the implementation of \textit{Overloaded} is .  
%
% Moreover, the products of the SPL systems can be automatically generated by a selection of features. 
%

A set of the selections of all the features in $\mathbb{F}$ defines a \textit{configuration}. Any non-empty subset of a configuration is called a \textit{partial configuration}.
A configuration specifies a single \textit{product}.
For example, configuration $c_1 = \{Empty = F,Weight = T, TwoThirdsFull = F, Overloaded = F\}$ specifies product $p_1$. 
A product is the composition of the implementation of all the enabled features, e.g., $p_1$ is composed of $\varphi(Base)$ and $\varphi(Weight)$. 

We denote the sets of all the possible valid configurations and all the corresponding products of $\mathfrak{S}$ by $\mathbb{C}$ and $\mathbb{P}$, respectively ($|\mathbb{C}|=|\mathbb{P}|$).
%
%
% For system $\mathfrak{S} = \langle\mathbb{S}, \mathbb{F}, \varphi \rangle$, the number of all possible configurations (products), $\|\mathbb{C}\|=\|\mathbb{P}\|$ can be up to $2^{\|\mathbb{F}\|}$. Exhaustively testing all the possible products is prohibitively expensive. 
%
In practice, a subset of $\mathbb{C}$, $C$ (the corresponding products $P \subset \mathbb{P}$), is sampled for testing and finding bugs.
%
% Unlike non-configurable code, bugs in an SPL system can be \textit{variable} in different products, which causes the failures in certain products. 
Unlike non-configurable code, bugs in SPL systems can be \textit{variable} and only cause the failures in certain products.
In Table \ref{Table:example_test_result}, the bug at line 31 only causes the failures in $p_6$ and $p_7$. This bug is a \textit{variability bug}. In the rest of this paper, we simply call an SPL system containing variability bugs a \textbf{buggy system}. In this work, we localize bugs at the statement-level~\cite{ieee1990ieee}, which is the granularity widely adopted in the existing studies~\cite{pearson2017evaluating, zhang2017boosting, abreu2007accuracy, wen2019historical}.

\begin{definition}{\textbf{(Variability Bug).}}
\label{Def:config_dependent_bug}
Given a buggy SPL system $\mathfrak{S}$ and a set of products of the system, $P$, which is sampled for testing, a variability bug is an incorrect code statement of $\mathfrak{S}$ that causes the unexpected behaviors (failures) in a set of products which is a non-empty strict subset of $P$. 
\end{definition}{}

% \begin{definition}{\textbf{(Variability Bug).}}
% \label{Def:config_dependent_bug}
% Given an SPL system $\mathfrak{S}$ and a set of products of the system, $P$, which is sampled for testing, $\mathfrak{S}$ contains variability bug(s) if there are one or more incorrect code statement(s) of $\mathfrak{S}$ that causes the unexpected behaviors (failures) in a set of products which is a non-empty strict subset of $P$. 
% \end{definition}{}

In other words, $\mathfrak{S}$ contains variability bugs if and only if $P$ is categorized into two separate non-empty sets based on their test results: the \textit{passing products} $P_P$ and the \textit{failing products} $P_F$ corresponding to the \textit{passing configurations} $C_P$ and the \textit{failing configurations} $C_F$, respectively. Every product in $P_P$ passes all its tests, while each product in $P_F$ fails at least one test.
Note that $P_P \cup P_F = P$ and $C_P \cup C_F = C$.
% Note that $P_P \cup P_F = P$, $P_P \cap P_F = \emptyset$, $C_P \cup C_F = C$, and $C_P \cap C_F = \emptyset$.
%
In Table \ref{Table:example_test_result}, $P_P = \{p_1, p_2,$ $p_3, p_4, p_5\}$, and $P_F = \{p_6, p_7\}$. 
%
%
%
% %
Besides the test results, the statements of every product executed by each test are also recorded to localize bugs. This execution information is called program spectra~\cite{harrold1998empirical}.

% \begin{definition}{\textbf{(Program Spectra).}}
% For a product $p \in P$, a program spectra, $T$, of $p$ is a set of the executed statements which are recorded during testing.
% % and $T = T_F \cup T_P$, where $T_F$ and $T_P$ are the sets of the statements executed during running the failed and passed tests, respectively.  
% \end{definition}{}

%

From the above basic concepts, the problem of variability fault localization is defined as follows:

\begin{definition}{\textbf{(Variability Bug Localization).}}
\label{Def:config_dependent_bug_localize}
Given 3-tuple $\langle \mathfrak{S}, P, R\rangle$, where:
\begin{itemize}
    \item $\mathfrak{S}=\langle\mathbb{S},  \mathbb{F}, \varphi \rangle$ is a system containing variability bugs.
    % \item $P$ is the set of sampled products, $P=P_P \cup P_F$, where $P_P$ and $P_F$ are the sets of passing and failing products of $\mathfrak{S}$
    % \item $R$ is the set of the program spectra of all the products in $P$
    
     \item $P = \{p_1, p_2, ..., p_n\}$ is the set of sampled products, $P=P_P \cup P_F$, where $P_P$ and $P_F$ are the sets of passing and failing products of $\mathfrak{S}$.
    \item $R = \{
    R_1, R_2, ..., R_n\}$ is the set of the program spectra of the products in $P$, where $R_i$ is the program spectra of $p_i$.
\end{itemize}
Variability Bug Localization is to output the list of the statements in $\mathfrak{S}$ ranked based on their suspiciousness to be buggy.
\end{definition}{}

\section{Feature Interaction}
\label{sec:interaction}
We introduce our feature interaction formulation used in this work to analyze the root cause of variability bugs.

\subsection{Feature Interaction Formulation}
Different kinds of feature interactions have been discussed in the literature~\cite{apel2013exploring, garvin2011feature, ase19prioritization, soares2018varxplorer}. In this work, we formulate feature interaction based on the impacts of a feature on other features.
Specially, for a set of features in a product, a feature can interact with the others in two ways: (i) directly impacting the others' implementation and (ii) indirectly impacting the others' behaviors via the statements which are impacted by \textit{all} of them.
For (i), there is control/data dependency between the implementation of these features. 
For example, in $p_5$, since the statement at line 26 ($s_{26}$) in $\varphi(TwoThirdsFull)$ is data-dependent on $s_9$ and $s_{15}$ in $\varphi(Weight)$, there is an interaction between $Weight$ and $TwoThirdsFull$ in $p_5$.
For (ii), there is at least one statement which is control/data dependent on a statement(s) of every feature in the set.
For instance, in $p_5$, $TwoThirdsFull$ and $Weight$ interact by both impacting the statement at line 31.
As a result, when these features are all on in a product, 
% the presence of each of them 
each of them will impact the others' behaviors. 
Without loss of generality, a statement can be considered to be impacted by that statement itself. Thus, for a set of enabled features $F$, in a product, there exists an interaction among these features if there is a statement $s$ which is impacted by the implementation of all the features in $F$, regardless of whether $s$ is used to implement these features or not.
Formally, we define \textit{impact function}, $\Omega$, to determine the impact of a statement in a product.

\begin{definition}\textbf{(Impact Function).}
\label{def:influence}
Given a system $\mathfrak{S}=\langle\mathbb{S}, \mathbb{F}, \varphi \rangle$, we define impact function as $\Omega: \mathbb{S} \times \mathbb{P} \to 2^\mathbb{S}$. 
Specially, $\Omega(s, p)$ refers to the set of the statements of $\mathfrak{S}$ which are impacted by statement $s$ in product $p$. For a statement $s'$ in product $p$, $s' \in \Omega(s, p)$ if $s'$ satisfies one of the following conditions:
    \begin{itemize}
        \item $s' = s$
        \item $s'$ is data/control-dependent on $s$
    \end{itemize}
\end{definition}{}
For our example, $\Omega(s_{33}, p_5) = \{s_{33}, s_{36}, s_{37}, s_{38}\}$.
Note that if statement $s$ is not in product $p$, then $\Omega(s, p) = \emptyset$.

\begin{definition}{\textbf{(Feature Interaction).}}
\label{Def:feature_interaction}
Given a system $\mathfrak{S}=\langle\mathbb{S},  \mathbb{F}, \varphi \rangle$, for product $p$ and a set of features $F$ which are enabled in $p$, the interaction among the features in $F$ exists if and only if the following condition is satisfied:
$$\bigcap_{f\in F} \alpha(f, p) \neq \emptyset$$
where $\alpha(f, p) = \bigcup_{s \in \varphi(f)} \Omega(s, p)$ refers to the set of the statements in $p$ which are impacted by any statement in $\varphi(f)$. 
The \textbf{implementation of the interaction} among features $F$ in product $p$ is denoted by $\beta(F, p) = \bigcap_{f\in F} \alpha(f, p)$.
\end{definition}{}

% In other words, all the features in $F$ interact with each other in $p$ when there is at least one statement impacted by all their implementation.
%

% \begin{definition}{\textbf{(Feature Interaction Implementation).}}
% \label{Def:feature_interaction_implementation}
% Given a set of features $F$, where the features in $F$ interact with each others in a product $p$, the implementation of this interaction is:
% %
% $$\beta(F, p) = \bigcap_{f\in F} \alpha(f, p)$$
% %
% \end{definition}{}

% In this work, the \textit{implementation of the interaction} among features $F$ in product $p$ is denoted by $\beta(F, p)$. Particularly,  $\beta(F, p) = \bigcap_{f\in F} \alpha(f, p)$.
For the example in Fig. \ref{fig:example_code}, the features $Weight$ and $Overloaded$ interact with each other in product $p_7$ and the implementation of the interaction includes the statements at lines 31--33 and 36--38. Note that, without loss of generality, a feature can impact and interact with itself.

%Check buggy statement concept
\subsection{The Root Cause of Variability Failures}
\label{sec:root_cause}

In this section, we analyze and discuss the relation between variability failures in SPL systems and the enabling/disabling of system features. 
In a buggy SPL system, the variability bugs can be revealed by set(s) of the \textbf{relevant} features which must be enabled and disabled \textbf{together} to make the bugs visible. 
For each set of relevant features, their selections might affect the visibility of the bugs in the system. 
For simplicity, we first analyze the buggy system containing a single set of such relevant features. The cases where multiple sets of relevant features involve in the variability failures will be discussed in the later part.

Let consider the cases where the failures of a system are revealed by \textit{a single set of the relevant features}, $F_r = F_E \cup F_D$, where the features in $F_E$ and $F_D$ must be respectively enabled and disabled \textbf{together} to make the bugs visible.
Specially, the features in $F_E$ and $F_D$ must be respectively on and off in all the failing products.
From a failing product $p \in P_F$, once switching the current selection of any
switchable 
feature\footnote{In configuration $c$, feature $f$ is switchable if switching $f$'s selection, the obtained configuration is valid regarding system's feature model.} %
%
%
%\footnote{A feature $f$ in configuration c is switchable if switching the selection of $f$, the obtained configuration is still valid regarding to the feature model. Without loss of generality, we assume all the features of the system are switchable.}
%
in $F_r$, the resulting product $p'$ will pass all its tests, $p' \in P_P$. 
%
% Intuitively, the features in $F_r$ must interact with each others.
% in the products where they are on.
%
In this case, the interaction of features in $F_r$ propagates the impact of buggy statements to the actual outputs causing the failures in $p$.
The relation between variability bugs \textit{(buggy statements)} causing the failures in failing products and the interaction between the relevant features in $F_r$ will be theoretically discussed as following.

For a failing product $p \in P_F$, we denote the set of the buggy statements in $p$ by $\mathcal{S}_b$.
From $p$, disabling any feature $f_e \in F_E$ would produce a passing product $p' \in P_P$. In this case, \textbf{every} buggy statement $s \in \mathcal{S}_b$ can be either present or not in $p'$. 
First, if $s$ is not in $p'$ after disabling $f_e$ from $p$, then $s \in \varphi(f_e) \subseteq \alpha(f_e, p)$.
The second case is that  $s$ is still in $p'$. Due to the absence of $f_e$,  $s$ \textit{behaves differently} from the way it does incorrectly in $p$, and $p'$ passes all its tests. 
This means, in $p$, $f_e$ has impact on $s$ and/or the statements impacted by $s$, $\Omega(s, p)$. In other words, $\alpha(f_e, p) \cap \Omega(s, p) \neq \emptyset$.
In $p$, $s$ and the statements impacted by $s$ together propagate their impacts to the unexpected outputs in $p$. Thus, any change on the statements in $\Omega(s, p)$ can affect the bugs' visibility.

These two above cases show that \textit{every} incorrect statement $s$ in $\mathcal{S}_b$, only exposes its bugginess with the presence of \textit{all} the features in $F_E$.
This demonstrates that the features in $F_E$ must interact with each other in $p$, $\beta(F_E, p) = \bigcap_{f_e\in F_E}\alpha(f_e, p) \neq \emptyset$.
Indeed, if there exists a feature $f \in F_E$ which does not interact with the others in $F_E$, $\alpha(f, p) \cap \beta(F_E \setminus \{f\}, p) = \emptyset$, the incorrect behaviors of $s$ will only be impacted by either $f$ or the interaction among the features in \{$F_E \setminus {f}\}$. As a result, $s$ will not require the presence of both $f$ and $\{F_E \setminus {f}\}$ to reveal its bugginess.
% In fact, if the impacts of the features in $F_E$ are independent, a buggy statement $s \in \mathcal{S}_b$ will not require all the features in $F_E$ enabled together to reveal the statement's bugginess.
%
% , where $\beta(F_E, p) = \bigcap_{f_e\in F_E} \alpha(f_e, p)$ is the implementation of the feature interaction among $F_E$ in $p$. 
%
%
Moreover, since every $f_e \in F_E$ has impacts on the behaviors of all the buggy statements,  $\forall s \in \mathcal{S}_b$, $\forall f_e \in F_E, \Omega(s, p) \cap \alpha(f_e, p) \neq \emptyset$, the interaction among the features in $F_E$ also has impacts on the behaviors of every buggy statement,
$\forall s \in \mathcal{S}_b$, $\Omega(s, p) \cap \beta(F_E, p) \neq \emptyset$.
In other words, 
the features $F_E$ interact with each other in $p$, and the interaction implementation impacts the visibility of the failures caused by every buggy statement.
Hence, the statements which implement the interaction among the features in $F_E$ are valuable suggestions to localize the buggy statements.
This theoretically confirms our observation \textbf{O1}.
%(Mentioned in \textbf{O1}).

% Similarly, from $p$, turning on any disabled feature $f_d \in F_D$, the resulting product $p''$ also passes all its tests. Because the bug-relevant statements of \textbf{every} buggy statement $s$ in $\mathcal{S}_b$ are impacted by the presence of $f_d$, the incorrect behaviors of $s$ cannot be exposed in $p''$.
% %
% Formally, $\mathcal{S}_r(s, p) \cap \alpha(f_d, p'') \neq \emptyset$, where $\mathcal{S}_r(s, p)$ refers to the bug-relevant statements of $s$ in $p$.
% %
% Intuitively, $f_d$ has impacts on the implementation of the interaction among $F_E$ in $p$ as well as the impact of this interaction on $\mathcal{S}_r(s, p)$.
% Hence, \textit{the interaction among $F_E \cup \{f_d\}$ impacts the bug-relevant statements of every buggy statement $s \in \mathcal{S}_b$, $\mathcal{S}_r(s, p) \cap \beta(F_E \cup \{f_d\}, p'') \neq \emptyset$}.
% This explains our observation \textbf{O2}.
% % (Mentioned in \textbf{O2}).

Similarly, from $p$, turning on any disabled feature $f_d \in F_D$, the resulting product $p''$ also passes all its tests.
This illustrates that in $p''$, the behaviors of \textbf{every} buggy statement $s$ in $\mathcal{S}_b$ are impacted by the presence of $f_d$, thus the incorrect behaviors of $s$ cannot be exposed.
Formally, $\forall s \in \mathcal{S}_b, \Omega(s, p) \cap \alpha(f_d, p'') \neq \emptyset$.
Intuitively, $f_d$ has impacts on interaction implementation of $F_E$ in $p$ as well as the impact of this interaction on $\Omega(s, p)$.
In other words, \textit{the interaction among $F_E \cup F_D$ impacts the behaviors of the buggy statements, $\forall s \in \mathcal{S}_b$, $\Omega(s, p) \cap \beta(F_E \cup F_D, p'') \neq \emptyset$}. Hence, investigating the interaction implementation of the features in $F_D$ and $F_E$ (if they were all enabled in a product) can provide us useful indications to find the incorrect statements.  
This explains our observation \textbf{O2}.
% (Mentioned in \textbf{O2}).
%NEW

% \textbf{CHECKING THIS PART...}

Overall, the interaction among the relevant features in $F_r = F_E \cup F_D$ reveals/hides the bugs by impacting the buggy statements and/or the statements impacted by the buggy ones in the products. 
Illustratively, this interaction implementation propagates the impact of all the buggy statements to the output of the failed tests in a failing product.
% From the implementation of the interaction among these relevant features, the buggy statements can be isolated by identifying all the statements which are impacted by or have impacts on the interaction implementation.
%
This means that the buggy statements are contained in the set of statements
which are impacted by or have impacts on the interaction implementation of the relevant features.
%
%
% Note that, for any statement $s$ in $\mathcal{S}_r$ or $\beta(F_E, p)$, all the statements impacted by $s$ are also consisted in that set. Thus, if a statement $s \in \mathcal{S}_r \cap \beta(F_E, p)$, all the statements impacted by $s$ in this product will also be included in the common part of $\mathcal{S}_r$ and $\beta(F_E, p)$. Intuitively, from the interaction implementation of features in $F_r = F_E \cup F_D$, by identifying all the statements which have impacts on this interaction, we can identify a super-set of $\mathcal{S}_r$. In addition, this identified set will contain buggy statements because $\mathcal{S}_b \subseteq \mathcal{S}_r$.
%
Thus, \textit{identifying the sets of the relevant features whose interaction can affect the variability bugs visible/invisible and the implementation of the interaction is necessary to localize variability bugs in SPL systems}.
%

% how would SBFL (with/out slicing) perform in this case? How would this performance be different from the proposed approach?
%
% For SBFL and its combination with the slicing method~\cite{?}, without considering this important relation between variability failures and feature selections, they might misleadingly treat correct statements as suspicious. 
% %
% %SBFL: all the statements executed during running failed tests (all the statements would be nice :D).
% %S-SBFL: point out a correct statement which is misleadingly isolated...
% In facts, the suspicious space of SFBL can be narrowed down when SFBL is combined with the slicing techniques (S-SBFL). However, S-SBFL still considers the statements which are unrelated to the fault, such as the statement at line 36 in Fig~\ref{fig:example_code}, as suspicious.
% %
% This problem becomes more severe in large SPL systems. This will be empirically illustrated in our experimental results (Section \ref{sec:empirical_results}).

In general, variability bugs in a system can be revealed by multiple sets of relevant features. In these cases, the visibility of the bugs might not be clearly observed by switching the selections of features in one set of relevant features. For instance, $F_r$ and $F_r'$ are two sets of relevant features whose interaction causes the failures in the system. Once switching the selection of any feature in $F_r$, the implementation of the interaction among the features in $F_r$ is not in the resulting product $p'$. Meanwhile, $p'$ can still contain the interaction among the features in $F_r'$. Thus, $p'$ can still fail some tests. 
However, if we can identify $F_r$ and/or $F_r'$ or even their subsets, by examining the interaction among the identified set(s) of features, the bugs can be isolated. More details will be described and proved in the next section.

In spite of the importance of the relation between variability failures and relevant features, this information is not utilized by existing studies such as SBFL and S-SBFL. 
Consequently, their resulting suspicious spaces are often large.
Meanwhile, by Arrieta et al.~\cite{arrieta2018spectrum}, SBFL is adapted to localize bugs at the feature-level. 
Particularly, each sampled product is considered as a test (i.e., passed tests are passing products, and failed tests are failing products), and the spectra record the feature selections in each product. However, SBFL is used to localize the \textit{buggy features}. By this method, all the statements in the same feature have the same suspiciousness level. Thus, this approach could be ineffective for localizing variability bugs at the statement-level.
This will be empirically illustrated in our results (Section \ref{sec:empirical_results}).

%TRANG
% In summary, in this paper, we introduce an approach which identifies and leverages the interaction among
% the relevant features to localize the bugs.
% Although identifying the relevant feature sets which affect bug visibility is significantly important, this information is not utilized by
% SBFL and its combination with the slicing method (S-SBFL). Consequently, the suspicious spaces resulted from these approaches are often large.
% Furthermore, there is another approach, proposed by Arrieta et al.~\cite{arrieta2018spectrum}, 
% which also adapts SBFL for fault localizing in SPL systems. They consider each sampled product as a test (i.e., passed tests are passing products and failed tests are failing products), and the spectra record the selections of the features in each product. However, by this method, SBFL  is used to localize the buggy features and all the statements in the same feature will have the same suspicious level. Therefore, this adaptation could be ineffective for localizing variability bugs at the statement level.
% This will be empirically illustrated in our experimental results (Section \ref{sec:empirical_results}).

\section{Buggy Partial Configuration Detection}
\label{sec:spc_detection}

In this section, we introduce the notions of \textit{Buggy Partial Configuration (\bpc)} and \textit{Suspicious Partial Configuration (\spc)}. 
Specially, \bpcs are the partial configurations whose interactions among the corresponding features are the root causes making variability bugs visible in a buggy system.
In general, \bpcs can be detected after testing all the possible products of the system. However, verifying all those products is nearly impossible in practice. 
Meanwhile, \spcs are the detected suspicious candidates for the \bpcs which can be practically computed using the sampled products.

\subsection{Buggy Partial Configuration}
\label{sec:bpc}

For a buggy system $\mathfrak{S}=\langle\mathbb{S}, \mathbb{F}, \varphi \rangle$, where all the possible configurations of $\mathfrak{S}$, $\mathbb{C}$, is categorized into the non-empty sets of passing ($\mathbb{C}_P$) and failing ($\mathbb{C}_F$) configurations, $\mathbb{C}_P \cup \mathbb{C}_F = \mathbb{C}$,
a Buggy \textbf{P}artial \textbf{C}onfiguration (\bpc) is the minimal set of feature selections that makes the bugs visible in the products. In Fig. \ref{fig:example_code}, the only \bpc is $B = \{Weight = T, TwoThirdsFull = F, Overloaded = T\}$. 

\begin{definition}{\textbf{(Buggy Partial Configuration (\bpc)).}}
\label{Def:bpc}
Given a buggy system $\mathfrak{S}=\langle\mathbb{S}, \mathbb{F}, \varphi \rangle$, a buggy partial configuration, $B$, is a set of feature selections in $\mathfrak{S}$ that has both the following Bug-Revelation and Minimality properties:
    \begin{itemize}
        \item \textbf{Bug-Revelation}. Any configuration containing $B$ is corresponding to a failing product, $\forall c \in \mathbb{C}, c \supseteq B \implies c \in \mathbb{C}_F$. 
        
        \item \textbf{Minimality}. There are no strict subsets of $B$ satisfying the Bug-Revelation property, $\forall B' \subsetneq B \implies \neg(\forall c \in \mathbb{C}, c \supseteq B' \implies c \in \mathbb{C}_F)$.
    \end{itemize}

\end{definition}{}

%Why BPC needs bug-revelation property?
\textbf{Bug-Revelation}. 
This property is equivalent to that \textit{all the passing configurations do not contain $B$}. Indeed,
$$\forall c \in \mathbb{C}, c \supseteq B \implies c \in \mathbb{C}_F$$ 
$$\Leftrightarrow \forall c \in \mathbb{C}, [c \not\supseteq B \vee c \in \mathbb{C}_F]$$ 
$$\Leftrightarrow \forall c \in \mathbb{C}, [c \not\supseteq B \vee c \not\in \mathbb{C}_P]$$ 
$$\Leftrightarrow \forall c \in \mathbb{C}, c \in \mathbb{C}_P \implies c \not \supseteq B$$
For a set of feature selections $B'$, if there exists a passing configuration containing $B'$, $\exists c \in \mathbb{C}_P \wedge c \supset B'$,
the interaction among the features in $B'$ in the corresponding product $p$ cannot be the root cause of any variability bug.
This is because there is no unexpected behavior caused by this interaction in the passing product $p$\footnote{Assuming that the test suite of each product is effective in detecting bugs. This means the buggy products must fail.}. 
Hence, investigating the interaction between them \textit{might not} help us localize the bugs.
For example, $B'=$ $\{Empty = T, Weight = T\}$ is a subset of the failing configuration $c_6$, however it is not considered as a \bpc, because $B'$ also is a subset of $c_2$ which is a passing configuration. Indeed, in every product, the interaction between $Empty$ and $Weight$, which does not cause any failure, should not be investigated to find the bug. 
Thus, to guarantee that the interaction among the features in a \bpc is the root cause of variability bugs, the set of feature selections needs to have \textit{Bug-Revelation} property.
%

%why BPC needs minimality property?
\textbf{Minimality}. If a set $B$ holds the \textit{Bug-Revelation} property but not \textit{minimal}, then there exists a strict subset $B'$ of $B$ ($B' \subsetneq B$) that also has \textit{Bug-Revelation} property.
%
%
%%%%THE SECOND CASE
%
However, for any $p \in P_F$ whose configuration contains both $B$ and $B'$, to detect all the bugs related to either $B$ or $B'$, the smaller one, $B'$, should still be examined rather than $B$. 
The reason is, in $p$, the bugs related to both $B$ and $B'$ are all covered by the implementation of the interaction among the features in $B'$. 
Particularly, let $B' = F'_{E} \cup F'_{D}$ and $B = F_{E} \cup F_{D}$, where $F'_{E}$, $F'_{D}$, $F_{E}$, and $F_{D}$ are the sets of enabled and disabled features in $B'$ and $B$, respectively. Since $B' \subset B$, and $F'_{E} \cap F'_{D}= F_{E} \cap F_{D}=\emptyset$, we have $F'_{E} \subseteq F_{E}$ and $F'_{D} \subseteq F_{D}$.
For the enabled features in $F_{E}$ and $F'_{E}$, the failures in $p$ can be caused by the interactions among the enabled features in $F_{E}$ or $F'_{E}$. The implementation of the interactions among the enabled features in $F_{E}$ and $F'_{E}$ in $p$ are $\beta(F_{E}, p) = \bigcap_{f \in F_{E}}\alpha(f, p)$ and $\beta(F'_{E}, p) = \bigcap_{f' \in F'_{E}}\alpha(f', p)$, respectively. 
Then, we have $\beta(F_{E}, p) \subseteq \beta(F'_{E}, p)$ because $F'_{E} \subseteq F_{E}$. 
As a result, the incorrect statements related to $\beta(F_{E}, p)$ are all included in $\beta(F'_{E}, p)$.
Similarly, for the sets of the disabled features, the set of the statements in $p$ related to all the features of $F'_D$ also includes the statements related to all the features of $F_{D}$.
In consequence, by identifying the interaction implementation of the features in $B'$, the bugs which are related to both $B$ and $B'$ can all be found. 

Furthermore, if both $B$ and $B'$ are related to the same bug(s), the larger set could contain bug-irrelevant feature selections which can negatively affect the FL effectiveness.
% Thus, to avoid missing the buggy statements related to both sets of feature selections $B$ and $B'$, the smaller one, $B'$, should be investigated instead of $B$. 
For example, both the entire configuration $c_6$ and its subset $\{ TwoThirdsFull = F, Overloaded = T, Weight = T\}$ has \textit{Bug-Revelation} property. Nevertheless, the interaction among $TwoThirdsFull$, $Overloaded$, and $Weight$, which indeed causes the failures in $p_6$, should be investigated instead of the interaction among the features in the entire $c_6$. 
This configuration contains the bug-irrelevant selection, $Empty = T$.
As a result, \bpcs need to be \textit{minimal}.

\textbf{\bpc Detection Requirement}. All possible configurations of a system are very rarely available for the QA process. 
Thus, in a buggy system, the sampled set  is used to detect  the \bpcs. 
Assuming that $\mathbb{D}$ is the set of the candidates for \bpcs  which is detected by an FL approach. 
For any $D\in \mathbb{D}$, all the statements which implement the interaction among the features in $D$ and the statements which have impact on that interaction implementation (Section~\ref{sec:root_cause}) are suspicious.
Thus, for a \bpc $B$, to avoid missing related buggy statements $\mathcal{S}$, the FL method must ensure that $\mathcal{S}$ is covered by the suspicious statements identified from least one candidate: \textit{there must be at least one candidate in $\mathbb{D}$ which is a subset of $B$, $\exists D \in \mathbb{D}$, $D \subseteq B$}. Let us call this the \textbf{effectiveness requirement}. 

Indeed, if there exists $D \in \mathbb{D}$, such that $D \subseteq B$, then in a product $p$ whose configuration contains $B$ (apparently contains $D$), the interaction implementation of the features in $D$ covers all the statements implementing the interaction of the features in $B$, $\beta(D,p) \supseteq \beta(B,p)$. 
% Therefore, $\beta(D,p)$ contains $\mathcal{S}$ which is included in $\beta(B,p)$. 
%
As a result, the suspicious statements set of $D$ contains both the interaction implementation $\beta(B,p)$ and the statements which have impact on this interaction in $p$. In other words, the suspicious statements set of $D$ contains buggy statements $\mathcal{S}$.
%
% If there is no candidate which is a subset of $B$, $\mathcal{S}$ might not be captured by the interaction of any candidate. Consequently, the FL technique is \textit{ineffective} in localizing the bugs.
Hence, to guarantee the effectiveness in localizing variability bugs, we aim to detect the set of the candidates for \bpcs which satisfies the effectiveness requirement.

\subsection{Important Properties to Detect \bpc}
\label{subsec:spc_properties}

In practice, a system $\mathfrak{S}$ may have a huge number of possible configurations, $\mathbb{C}$. Consequently, only a subset of $\mathbb{C}$ is sampled for testing and debugging,
% In practice, all the possible configurations of a system $\mathfrak{S}$, $\mathbb{C}$, is rarely available for the QA process. Instead, a subset of $\mathbb{C}$ is sampled for testing and debugging,
$C= C_P \cup C_F$.
A set of feature selections which has both
\textit{Bug-Revelation} and \textit{Minimality} properties on the sampled set $C$ is intuitively \textit{suspicious} to be a \bpc. 
Let us call these sets of selections \textit{Suspicious Partial Configurations (\spcs)}.

%example SPC
For example, in Fig. \ref{fig:example_code}, $D = \{TwoThirdsFull = F, Overloaded = T\}$ is a \spc. All the configurations containing $D$ ($c_6$ and $c_7$) are failing. Additionally, all the strict subsets of $D$ do not hold \textit{Bug-Revelation} on $C$, e.g., $\{TwoThirdsFull = F\}$ is in $c_1$, and $\{Overloaded = T\}$ is in $c_5$, which are passing configurations. 
%
% Thus, $D$ is minimal on $C$. 
%
Thus, $D$ is a minimal set which holds the \textit{Bug-Revelation} property on $C$.
%
%
%
% Note that \spcs and \bpcs all hold the \textit{Bug-Revelation} on the given set of configurations, $C$. However, because there are some (passing) configurations in $\mathbb{C}$ but not in $C$, it does not express that some selections must be in a \bpc. 
% Hence, \textit{\bpcs might not hold \textit{Minimality} property on $C$}.
%
% In Fig.~\ref{fig:example_code}, a \bpc is $B = \{Weight = T, TwoThirdsFull = F, Overloaded = T\}$. However, $B$ does not satisfy \textit{Minimality} on the set of the available configurations (Table~\ref{Table:example_test_result}), because its subset, $\{TwoThirdsFull = F, Overloaded = T\}$, satisfies \textit{Bug-Revelation} on $C$.
%
% The reason is that Table~\ref{Table:example_test_result} does not show any product whose configuration contains both $\{Weight = F\}$, as well as, $\{TwoThirdsFull = F, Overloaded = T\}$ passes or fails their tests. Therefore, $\{Weight = T\}$ is not expressed as a part of the \bpc.
%
% The reason is, $\{Weight = T\}$ is not shown as a part of \bpc. Since, in Table~\ref{Table:example_test_result}, the test information does not show that whether the product whose configuration contains $\{Weight = F, TwoThirdsFull = F, Overloaded = T\}$ passes or fails its tests.

%
%
% Intuitively, every failing configuration $c$ contains at least one \bpc. 
%
Theoretically, the \spcs in $c$ can be detected by examining all of its subsets to identify the sets satisfying both \textit{Bug-Revelation} and \textit{Minimality}. The number of sets that need to be examined could be $2^{|c|}-1$. However, not every selection in $c$ participates in \spcs. Hence, to detect \spcs efficiently, we aim to identify a set of the selections, $SFS_c$, (\textit{Suspicious Feature Selections}) of $c$ in which the selections potentially participate in one or more \spcs. Then, instead of inefficiently examining all the possible subsets of $c$, the subsets of $SFS_c$, which is a subset of $c$, are inspected to identify \spcs.

Particularly, in failing configuration $c$, there exist the selections such that switching their current states (from \textit{on} to \textit{off}, or vice versa) results in a passing configuration $c'$. In other words, the bugs in the product of $c$ are invisible in the product corresponding to $c'$, and the resulting product passes all its tests. 
% In other words, after switching these selections, the resulting product passes. 
Intuitively, each of these selections might be relevant to the visibility of the bugs. Each of these selections can be considered as a \textit{Suspicious Feature Selection}.
Thus, a selection, which is in a failing configuration yet not in a passing one, is suspicious to the visibility of the bugs.

\begin{definition}{\textbf{(Suspicious Feature Selection (SFS)).}}
For a failing configuration $c \in C_F$, a feature selection $f_s \in c$ is suspicious if $f_s$ is not present in at least one passing configuration, formally $\exists c'\in C_P, f_s \not\in c'$, specially, $f_s \in c\setminus c'$.
\end{definition}

For example, the \textit{$SFS_7$} of $c_7$ contains the selections in the set differences of $c_7$ and the passing configurations, e.g., $c_7 \setminus c_1 = \{Overloaded = T\}$. Intuitively, the set difference $(c_7 \setminus c_1)$ must contain a part of every \bpc in $c_7$, otherwise $c_1$ would contain a \bpc and fail some tests.
Hence, for any failing configuration $c$, we have the following property about the relation between the set of the \bpcs in $c$ and $(c \setminus c')$ where $c'$ is a passing configuration.

\begin{property}
\label{prop:suspicious_selections}
Given failing configuration $c$ whose set of the \bpcs is $BPC_c$, the difference of $c$ from any passing configuration $c'$ contains a part of every \bpc in $BPC_c$. Formally, $\forall c' \in C_P, \forall B \in BPC_c, (c \setminus c') \cap B \neq \emptyset$.
\end{property}
The intuition is that for a failing configuration $c$ and a passing one $c'$, a \bpc in $c$ can be either in their common part ($c \cap c'$) or their difference part ($c \setminus c'$). As $c'$ is a passing configuration, the difference must contain a part of every \bpc in $c$. Otherwise, there would exist a \bpc in both $c$ and $c'$, and  $c'$ should be a \textit{failing} configuration (\textit{Bug-Revelation} property). This is impossible.

%%%%PROVING THERE IS A SUBSET OF SFS_c WHICH SATISFIES BUG_REVELATION CONDITION

For a configuration $c \in C_F$, we denote $SFS_c$ as the set of all the suspicious feature selections of $c$, $SFS_c = \bigcup_{c' \in C_P} (c \setminus c')$. 
To detect \bpcs in $c$, we identify all subsets of $SFS_c$ which satisfy the \textit{Bug-Revelation} and \textit{Minimality} properties regarding to $C$.
% in $C$ is suspicious to be a \bpc of $c$. 
%
The following property demonstrates that our method maintains the \textit{effectiveness requirement} in detecting \bpcs (Section~\ref{sec:bpc}).

\begin{property}
\label{prop:sfs}
Given a failing configuration $c$ whose set of the \bpcs is $BPC_c$, for any $B\in BPC_c$, there exists a subset of $SFS_c$, $M \subseteq SFS_c$, such that $M$ satisfies the \textit{Bug-Revelation} condition in the sampled set $C$ and $M \subseteq B$.
\end{property}

\begin{proof}{}
Considering a \bpc of a failing configuration $c$, $B \in BPC_c$, we denote $(c \setminus c') \cap B$ by $M_{c'} \neq \emptyset$, where $c'$ is a passing configuration (Property \ref{prop:suspicious_selections}). Let consider $M = \bigcup_{c' \in C_P} M_{c'}$. Note that, as $M_{c'} \subseteq (c \setminus c')$, we have $M \subseteq SFS_c = \bigcup_{c' \in C_P} (c \setminus c')$. In addition, since $M_{c'} \subset B$ for every passing configuration $c' \in C_P$, we have $M \subset B$. Moreover, for every passing configuration $c' \in C_P$, because $c' \not\supset M_{c'}$ (since $M_{c'} \subset (c \setminus c')$), $c'$ does not contain any superset of $M_c$, then $c' \not\supset M$. As a result, $M$, which is a subset of both $SFS_c$ and $B$, satisfies \textit{Bug-Revelation} property.

\end{proof}

As a result, given $c \in C_F$, there always exists a common subset of $SFS_c$ and $B$, for any $B \in BPC_c$, and that set satisfies \textit{Bug-Revelation} property on $C$.
Hence, according to the \textit{effectiveness requirement} (Section  \ref{sec:bpc}), detecting \bpcs by examining the subsets of $SFS_c$ is effective to localize the variability bugs.
Furthermore, as $SFS_c$ only contains the differences of $c$ and other passing configurations, $|SFS_c| \leq |c|$. 
For example in Table \ref{Table:example_test_result}, \textit{SFSs} of configuration $c_6$ is $SFS_6 = \{Empty = T, Weight = T, TwoThirdsFull = F, Overloaded = T\}$, so $|SFS_6| < |c_6|$. Thus, $SFS_c$ should be used to detect \bpcs rather than $c$.

Note that \spcs and \bpcs all hold the \textit{Bug-Revelation} on the given sampled set of configurations, $C$. However, because there are some (passing) configurations in $\mathbb{C}$ but not in $C$, it does not express that some selections must be in a \bpc. 
Hence, \textit{\bpcs might not hold \textit{Minimality} property on $C$}.
In Fig.~\ref{fig:example_code}, a \bpc is $B = \{Weight = T, TwoThirdsFull = F, Overloaded = T\}$. However, $B$ does not satisfy \textit{Minimality} on the set of the available configurations (Table~\ref{Table:example_test_result}), because its subset, $\{TwoThirdsFull = F, Overloaded = T\}$, satisfies \textit{Bug-Revelation} on $C$.
The reason is that Table~\ref{Table:example_test_result} does not show any product whose configuration contains both $\{Weight = F\}$ as well as $\{TwoThirdsFull = F, Overloaded = T\}$ passes or fails their tests. Therefore, $\{Weight = T\}$ is not expressed as a part of the \bpc.

\subsection{\bpc Detection Algorithm}
\label{sec:spc_detection}

Algorithm \ref{alg:spc_detection} describes our algorithm to detect \bpcs and return the \spcs in a buggy system, given the sets of the passing and failing configurations, $C_P$ and $C_F$. 
% Algorithm \ref{alg:spc_detection} describes our algorithm to detect \spcs in a buggy system, given the sets of the passing and failing sampled configurations, $C_P$ and $C_F$. 

\begin{algorithm}
  \SetAlgoLined\DontPrintSemicolon
  \SetAlgoLined\DontPrintSemicolon
  \SetKwFunction{proc}{DetectBuggyPCs}
  
  \SetKwProg{DetectBuggyPCs}{Procedure}{}{}
  \DetectBuggyPCs{\proc{$C_P$, $C_F$}}{
  $SuspiciousPCSets = \emptyset$ \;
  \For{$c \in C_F$}{
    $SFS_c = \emptyset$ \;
    \For{$c' \in C_P$}{
        $F_S = c \setminus c'$ \;
        $SFS_c = SFS_c \cup F_S$\;
    }
    \For {$k \in [1, K]$}{
        % \nl $newestSPCs = \emptyset$ \;
        $S_{k} = subsetWithSize(SFS_c, k)$ \;
        \For {($cand \in S_{k}) \wedge (\nexists spc \in SuspiciousPCSets, cand \supset spc)$}{ 
            \If{ $satisfy(cand, C_P, C_F)$}{
                $SuspiciousPCSets.add(cand)$ \;
                % \nl $newestSPCs.add(spc)$ \;
            }
        }
            % \nl \For{$spc \in newestSPCs$}{
            %     \nl \If { $spc \subset SFS_c$}{
            %         \nl $SFS_c.remove(spc)$ \;
            %     }
            % }
            
    }
 }
 
 \KwRet $SuspiciousPCSets$\;}
  
  \caption{\bpc Detection Algorithm}
  \label{alg:spc_detection}
\end{algorithm}

In Algorithm \ref{alg:spc_detection}, all the \spcs in the system are collected from the \spcs identified in each failing configuration $c$ (lines 3--17). 
From lines 4--8, the set of the suspicious selections in $c$ is computed. In order to do that, the differences of $c$ from all the passing configurations are gathered and stored in $SFS_c$ (lines 5--8).
%
% Note that, there might exist the case when both $S$ and $S'$ are in $\wp$ and $S' \subset S$. In this case, $S$ might contain irrelevant selections.
% % that are  to the bugs. 
% %(related to $S'$)å
% Thus, to achieve \textit{Minimality}, we keep in $\wp$ the sets of selections that have no subset (line 7) before computing $SFS_c$ at line 8.

Next, the \spcs in $c$ are the subsets of $SFS_c$ which have both \textit{Bug-Revelation} and \textit{Minimality} with respect to $C=C_F \cup C_P$ (lines 12--14). Each candidate, a set of feature selections $cand$, is checked against these properties by $satisfy$ (line 12).  In Algorithm \ref{alg:spc_detection}, the examined subsets of $SFS_c$ have the maximum size of $K$ (lines 9--10).
In other words, the considered interactions are up to $K$-way.
In practice, most of the bugs are caused by the interactions of \textit{fewer than 6 features}~\cite{garvin2011feature, kuhn2004software}. Thus, one should set $K = \textit{7}$ to ensure the efficiency.
Specially, the function $subsetWithSize(SFS_c, k)$ (line 10) returns all the subsets size $k$ of $SFS_c$.
Note that if a set is already a \spc, then any superset of it would not be a \spc (violating \textit{Minimality}). Thus, all the supersets of the identified \spcs can be early eliminated (line 11).
%
% When a set of feature selections S is an SPC, then any supersets of S cannot be an SPC, because they will not satisfy minimality condition. Therefore, we eliminate all detected SPCs from $\partial$ before considering a larger subset of $\partial$ (line 15-17).

In our example, from Table \ref{Table:example_test_result}, the detected \spcs are two sets $D_1 = \{\textit{TwoThirdsFull} = F, \textit{Overloaded} = T\}$ and $D_2 = \{\textit{Empty} = T, \textit{Overloaded} = T\}$.

\section{Suspicious Statements Identification}
For a buggy SPL system, the incorrect statements can be found by examining the statements which implement interactions of \bpcs as well as the statements impacting that implementation, as discussed in Section~\ref{sec:root_cause}. Thus, all the statements which implement the interactions of \spcs and the statements impacting them, are considered as suspicious to be buggy. 
In a product $p \in P_F$, for a \spc $D$ whose the sets of enabled and disabled features are $F_E$ and $F_D$, respectively.
% the visibility of the bugs in $p$ is affected by not only $F_E$ but also $F_D$. 
Hence, in $p$, the interaction implementation of $D$ includes the statements implementing the interaction of $F_E$ which can be impacted by the features in $F_D$ (if the disabled features were on in $p$).
% For a buggy SPL system, the suspicious statements are all the statements implementing the interactions of \spcs in the failing products. In a product $p \in P_F$, for a \spc whose the sets of enabled and disabled features are $F_E$ and $F_D$ respectively, the visibility of the bugs in $p$ is affected by not only the features in $F_E$ but also the features in $F_D$. Hence, in $p$, the suspicious statements are the statements implementing the interaction of $F_E$ which can be impacted by the features in $F_D$ (if the implementation of the disabled features were on in $p$). Intuitively, the statements in $p$ that impact or are impacted by these suspicious statements can also be considered to be suspicious.
%
% Additionally, we can narrow down the suspicious space by considering only the statements executed during running the failed tests of $p$.
%
% , $\beta(F_E, p)$  (Definition~\ref{Def:feature_interaction_implementation})

% However, the implementation of the features in $F_D$ is not present in $p$. Moreover, the features in $F_D$ can be mutually exclusive with other features enabled in $p$, which is constrained by the feature model~\cite{SPLBook}.
%
In practice, the features in $F_D$ can be mutually exclusive with other features enabled in $p$, which is constrained by the feature model~\cite{SPLBook}. Thus, the impact of $F_D$ on the implementation of the interaction of $F_E$ in $p$, $\beta(F_E, p)$ 
% (Definition~\ref{Def:feature_interaction_implementation})
might not be easily identified via the control/data dependency in $p$.
% or in other products where the features in $F_D$ and $F_E$ are all enabled. 
%
%
%
In this work, the impacts of the features in $F_D$ on statements in $p$ are approximately identified by using \textit{def-use} relationships of the variables and methods that are shared between the features in $F_D$ and $p$~\cite{ase19prioritization}. 
Formally, for a statement $s$, we denote $def(s)$ and $use(s)$ to refer to the sets of variables/methods \textit{defined} and \textit{used} by $s$, respectively.

\begin{definition}{\textbf{(Def-Use Impact).}}
\label{Def:disabled_feature_impact}
Given an SPL system $\mathfrak{S}=\langle\mathbb{S}, \mathbb{F}, \varphi \rangle$ and its sets of products $\mathbb{P}$, we define the \textit{def-use impact} function as $\gamma$: $\mathbb{F} \times \mathbb{P} \to 2^\mathbb{S}$, where $\gamma(f, p)$ refers to the set of the statements in $p$ which are impacted by any statement in the implementation of feature $f$, $\varphi(f)$, via the variables/methods shared between $\varphi(f)$ and $p$. Formally, for a statement $s$ in product $p$, $s \in \gamma(f, p)$ if one of the following conditions is satisfied:
\begin{itemize}
    \item $\exists t \in \varphi(f)$, $ def(t) \cap use(s) \neq \emptyset$
    \item s is data/control-dependent on s', and $s' \in \gamma(f, p)$  
\end{itemize}

\end{definition}{}

In summary, in a product $p \in P_F$, for a \spc whose the sets of enabled and disabled features are $F_E$ and $F_D$, the suspicious statements satisfy the following conditions: 
\textbf{(i)} implementing the interaction of the features in $F_E$ and $F_D$ or impacting this implementation;
\textbf{(ii)} executing the failed tests of $p$. 
% Specially, the interaction implementation of the features in $F_E$ and $F_D$ in $p$ are the statements in this product which implement the interaction of the enabled features in $F_E$ and potentially impacted by the presence of the disabled features in $F_D$.
For a buggy system, the suspicious space contains all the suspicious statements detected for all \spcs in every failing product.

\section{Suspicious Statements Ranking}
\label{sec:ranking}

To rank the isolated suspicious statements of a buggy system, {\tool} assigns a score to each of these statements based on the program spectra of the sampled products.
In \tool, the suspiciousness of each statement is assessed based on two criteria/dimensions: \textit{Product-based Assessment} and \textit{Test Case-based Assessment}.

% The first criterion is based on the overall test results of the products containing the statement (\textit{Product-based Assessment}). Meanwhile, the second one is assessed based on the detailed results of the tests executed by the statement (\textit{Test Case-based Assessment}). 
%
%
\subsection{Product-based Suspiciousness Assessment}
This criterion is based on the overall test results of the products containing the statement. 
Specially, in a buggy system, a suspicious statement $s$ could be executed in not only the failing products but also the passing products. Hence, from the product-based perspective, \textit{the (dis)appearances of $s$ in the failing and passing products could be taken into account to assess the statement's suspiciousness in the whole system}. 
% In this work, the appearances of the statements in the passing and failing products, $h(s, M)$, is used as the first ranking criterion. 
%
In general, the product-based suspiciousness assessment for $s$ could be derived based on the numbers of the passing and failing products where $s$ is contained or not. Intuitively, the more failing products and the fewer passing products where $s$ is contained, the more suspicious $s$ tend to be.
This is similar to the idea of SBFL when considering each product as a test.
Thus, in this work, we adopt SBFL metrics to determine the product-based suspiciousness assessment for $s$. Specially, for a particular SBFL metric $M$, the value of this assessment is determined by $ps(s, M)$ which adopts the formula of $M$ with the numbers of the passing and failing products containing and not containing $s$ as the numbers of passed and failed tests executed or not executed by $s$.

% reflects the suspiciousness of $s$ pervasively in all the products containing $s$. Specially, this ranking criterion can be estimated by using an appropriate SBFL metric $M$ based on the numbers of passing and failing products containing $s$. 

\subsection{Test Case-based Suspiciousness Assessment}
The test case-based suspiciousness of a statement $s$ is evaluated based on the detailed results of the tests executed by $s$.
Particularly, in each failing product containing $s$, the statement is \textit{locally} assessed based on the program spectra of the product. 
Then, the local scores of $s$ in the failing products are aggregated to form a single value which reflects the test case-based suspiciousness of $s$ in the whole system.

Particularly, the \textit{local test case-based suspiciousness} of statement $s$ can be calculated by using the existing FL techniques such as SBFL~\cite{pearson2017evaluating, keller2017critical, naish2011model, abreu2009spectrum, abreu2007accuracy}. In this work, we use a ranking metric of SBFL, which is the state-of-the-art FL technique, to measure the local test case-based suspiciousness of $s$ in a failing product.
Next, for a metric $M$, the aggregated test case-based suspiciousness of $s$, $ts(s, M)$, can be calculated based on the local scores of $s$ in all the failing products containing it.
%
%
% $$ts(s, M) = \sigma[local(s, p_1, M),  ..., local(s, p_k, M)]$$
%
% where $\sigma$ is an aggregation function which combines the local test case-based suspiciousness scores of $s$ in all the products in $P_F$, while $local(s, p_i, M)$ refers to the local value of $s$ in $p_i$ which is assigned by using metric $M$. 
In general, we can use any aggregation formula~\cite{bian1999comparing}, such as \textit{arithmetic mean}, \textit{geometric mean}, \textit{maximum}, \textit{minimum}, and \textit{median} to aggregate the local scores of $s$. 

However, the local test case-based scores of a statement, which are measured in different products, should not be directly aggregated.
The reason is that the scores of the statement in different products might be incomparable. 
Indeed, with some ranking metrics such as Op2~\cite{ding2013fault} or Dstar~\cite{wong2013dstar}, once the numbers of tests of the products are different, the local scores of the statements in these products might be in significantly different ranges. Intuitively, if these local scores are directly aggregated, the products which have larger ranges will have more influence on the suspiciousness score of the statement in the whole system. Meanwhile, such larger-score-range products are not necessarily more important in measuring the overall test case-based suspiciousness of the statement. Directly aggregating these incomparable local scores of the statement can result in an inaccurate suspiciousness assessment.
Thus, to avoid this problem, 
these local scores in each product should be \textit{normalized} into a common scale, e.g., $[0, 1]$, before being aggregated.
We will show the impact of the normalization as well as choosing the aggregation function and ranking metric on \tool's performance in Section \ref{sec:intrisic}.

\subsection{Assessment Combination}
Finally, the two assessment scores $ps(s,M)$ and $ts(s, M)$ of statement $s$ are combined with a \textit{combination weight}, $w \in [0, 1]$ to form a single suspiciousness score of the statement:
$$score(s, M) = w*ps(s, M) + (1-w)*ts(s, M)$$ 
Note that, to avoid the bias caused by the range-difference between the two criteria, these two scores should be normalized into a common range, e.g., $[0, 1]$ before the interpolation.
In the ranking process, the isolated suspicious statements are ranked according to their interpolated suspiciousness score $score(s, M)$. The impact of the combination weight $w$ on \tool's fault localization performance will be empirically shown in Section~\ref{sec:intrisic}.

\section{Empirical Methodology}
To evaluate our variability fault localization approach,  we seek to answer the following research questions:

\noindent\textbf{RQ1: \textit{Accuracy and Comparison}.} How accurate is {\tool} in localizing variability bugs? And how is it compared to the state-of-the-art approaches~\cite{arrieta2018spectrum, keller2017critical, wong2016survey}?

\noindent\textbf{RQ2: \textit{Intrinsic Analysis}.} How do the components including the suspicious statement isolation, the normalization, the suspiciousness aggregation function, and the combination weight contribute to {\tool}'s performance?

\noindent\textbf{RQ3: \textit{Sensitivity Analysis}.} How do various factors affect {\tool}'s performance including the size of sampled product set and the size of test suite in each sampled product?

\noindent\textbf{RQ4: \textit{Performance in Localizing Multiple Bugs}.} How does {\tool} perform on localizing multiple variability bugs?

\noindent\textbf{RQ5: \textit{Time Complexity}.} What is  {\tool}'s running time?

\subsection{Dataset}
\label{sec:data}
To evaluate \tool, we conducted several experiments on a large public dataset of variability bugs~\cite{ourdataset}. This dataset includes 1,570 buggy versions with their corresponding tests of six Java SPL systems which are widely used in SPL studies. There are 338 cases of a single-bug, and 1,232 cases of multiple-bug. 
% In addition, these variability bugs vary in mutation operators (i.e., which are used to create them), code elements and the number of involving features. 
The details are shown in Table~\ref{Table:dataset}.

In the benchmark~\cite{ourdataset}, to generate a large number of variability bugs, the bug generation process includes three main steps: Product Sampling and Test Generating, Bug Seeding, and Variability Bug Verifying. First, for an SPL system, a set of products is systematically sampled by the existing techniques~\cite{sampling_comparision}. To inject a fault into the system, a random modification is applied to the system's original source code by using a mutation operator. Finally, each generated bug is verified against the condition in Definition \ref{Def:config_dependent_bug} to ensure that the fault is a variability bug and caught by the tests. The detailed design decisions can be found in~\cite{ourdataset}.

To the best of our knowledge, this is the only public dataset containing the versions of the SPL systems that failed by variability bugs found through testing~\cite{ourdataset}. 
Indeed, Arrieta et al.~\cite{arrieta2018spectrum} also constructed a set of artificial bugs to evaluate their approach in localizing bugs in SPL systems. However, this dataset has not been public. 
Besides, Abal et al.~\cite{98bugs} and Mordahl et al.~\cite{mordahl2019empirical} also published their datasets of real-world variability bugs. 
%
% However, most of these bugs are compile-time bugs, and all of these bugs are not provided along with corresponding test suites. 
%
However, these datasets are not fit our evaluation well because all of these bugs are not provided along with corresponding test suites, and in fact, most of these bugs are compile-time bugs.
%
%
% Specially, Abal et al.~\cite{98bugs} investigated commit reports and Mordahl~\cite{mordahl2019empirical} used static analysis tools. Then the bugs were manually examined and simplified to construct the datasets. Therefore, they are not suitable for evaluating our approach.  
%
Specially, these variability bugs are collected by (manually) investigating bug reports~\cite{98bugs} or applying static analysis tools~\cite{mordahl2019empirical}.  

Before running \tool on the dataset proposed by Ngo et al.~\cite{ourdataset}, we performed a simple inspection for each case on whether the failures of the system are possibly caused by non-variability bugs. Naturally, there are bugs which may be classified as ``variability" because of the low-quality test suites which cannot reveal the bugs in some buggy products.
We found that there are 53/1,570 cases (19 single-bug cases and 34 multiple-bug cases) where among the sampled products in each case, the product  containing only the base feature and disabling all of the optional features fails several tests. These cases possibly contain non-variability bugs. We will discuss them in Section \ref{sec:intrisic}.

\begin{table}[]
\centering
\caption{Dataset Statistics}
\label{Table:dataset}
\begin{threeparttable}
\begin{tabular}{|l|c|c|c|c|c|c|}
\hline
\multirow{2}{*}{\textbf{System}} & \multicolumn{2}{c|}{\textbf{Details}} & \multicolumn{2}{c|}{\textbf{Test info}} & \multicolumn{2}{c|}{\textbf{
Bug info}} \\ \cline{2-7} 
                        & \textit{\#LOC}          & \textit{\#F}         & \textit{\#P  }         & \textit{Cov}            & \textit{\#SB}    & \textit{\#MB}   \\ \hline
ZipMe                   & 3460           & 13          & 25            & 42.9           & 55              & 249         \\ \hline
GPL                     & 1944           & 27          & 99            & 99.4           & 105             & 267         \\ \hline
Elevator-FH-JML         & 854            & 6           & 18            & 92.9           & 20              & 102         \\ \hline
ExamDB                  & 513            & 8           & 8             & 99.5           & 49              & 214         \\ \hline
Email-FH-JML            & 439            & 9           & 27            & 97.7           & 36              & 90          \\ \hline
BankAccountTP           & 143            & 8           & 34            & 99.9           & 73              & 310         \\ \hline
\end{tabular}
\begin{tablenotes}
    \item \textit{\#F }and \textit{\#P}: Numbers of features and sampled products.
    \item \textit{Cov}: Statement coverage (\%).
    \item \textit{\#SB} and \textit{\#MB}: Numbers of single-bug and multiple-bug cases.
\end{tablenotes}
\end{threeparttable}
\end{table}

\subsection{Evaluation Setup, Procedure, and Metrics}

\subsubsection{Empirical Procedure}

\textbf{Comparative Study.} 
For each buggy version, we compare the performance in ranking buggy statements of \tool, Arrieta et al.~\cite{arrieta2018spectrum}, SBFL~\cite{pearson2017evaluating, keller2017critical, naish2011model, abreu2009spectrum, abreu2007accuracy}, and the combination of slicing method and SBFL (\textit{S-SBFL})~\cite{chaleshtari2020smbfl, li2020more} accordingly. 
For SBFL, each SPL system is considered as a non-configurable code. SBFL ranks all the statements executed by failed tests. For S-SBFL, to improve SBFL, S-SBFL isolates all the executed failure-related statements in every failing product by slicing techniques~\cite{static_slicing} before ranking.
A failure-related statement is a statement included in at least a program slice which is backward sliced from a failure point in a failing product. 
In this experiment, we used 30 most popular SBFL metrics~\cite{keller2017critical, naish2011model, pearson2017evaluating}.
For each metric, $M$, we compared the performance of all 4 techniques using $M$.
%

% Note that, the performance of SBFL might be improved in localizing faults in the non-configurable programs by combining SBFL and slicing techniques~\cite{chaleshtari2020smbfl, li2020more}. However, for variability bugs, this combination is non-trivial. 
% The reason is, the slicing technique are conducted based on the control and data dependence of the statements~\cite{static_slicing, dynamic_slicing, wong2016survey}. For an SPL system, a statement can be included in multiple products, and the statement's control/data dependencies in different products are not uniform. Realizing this non-trivial combination for localizing variability bugs in SPL systems requires a serious theoretical study.
% %
% Thus, we compare \tool against only SBFL.

\textbf{Intrinsic Analysis.} 
We studied the impacts of the following components: \textit{Suspicious Statement Isolation}, \textit{Ranking Metric}, \textit{Normalization}, \textit{Aggregation Function}, and \textit{Combination Weight}.  We created different variants of \tool with different combinations and measured their performance.

\textbf{Sensitivity Analysis.} 
We studied the impacts of the following factors on the performance of \tool: \textit{Sample size} and \textit{Test set size}. 
% We varied them and measured performance. 
To systematically vary these factors, the sample size is varied based on $k$-wise coverage~\cite{kwise2012} and the numbers of tests are gradually added.

\subsubsection{Metrics} 
% This section introduces 3 metrics that we use to compare performance of fault localization approaches for variability bugs. 
We adopted \textit{Rank}, \textit{EXAM}~\cite{wong2008crosstab}, and \textit{Hit@X}~\cite{lo2014fusion} which are widely used in evaluating FL techniques~\cite{keller2017critical, pearson2017evaluating, arrieta2018spectrum}. We additionally applied \textit{Proportion of Bugs Localized (PBL)}~\cite{keller2017critical} for the cases of multiple variability bugs.

\noindent\textbf{Rank.} 
The lower rank of buggy statements, the more effective approach. 
If there are multiple statements having the same score, buggy statements are ranked last among them.
Moreover, for the cases of multiple bugs, we measured \textit{Rank}s of the first buggy statement (\textit{best rank}) in the lists.

\noindent\textbf{EXAM.}
\textit{EXAM}~\cite{wong2008crosstab} is the proportion of the statements being examined until the first faulty statement is reached:  
$$EXAM = \frac{r}{N} \times 100\%$$
where $r$ is the position of the buggy statement in the ranked list and $N$ is the total number of statements in the list. The lower \textit{EXAM}, the better FL technique.

\noindent\textbf{Hit@X.} \textit{Hit@X}~\cite{lo2014fusion} counts the number of bugs which can be found after investing \textit{X} ranked statements, e.g., \textit{Hit@1} counts the number of buggy statements correctly ranked $1^{st}$ among the experimental cases. 
In practice, developers only investigate a small number of ranked statements before giving up\cite{parnin2011automated}. Thus, we focus on $X \in [1, 5]$.

\noindent\textbf{Proportion of Bugs Localized (PBL). } \textit{PBL}~\cite{keller2017critical} is the proportion of the bugs detected after examining a certain number of the statements. The higher \textit{PBL}, the better approach.

\section{Empirical Results}
\label{sec:empirical_results}

\subsection{Performance Comparison (RQ1)}

Table \ref{Table:comparison} shows the average performance of \tool, SBFL, the combination of slicing method and SBFL (\textbf{S-SBFL}), and the feature-based approach proposed by Arrieta et al.~\cite{arrieta2018spectrum} (\textbf{FB}) on 338 buggy versions containing a single bug each~\cite{ourdataset} in \textit{Rank} and \textit{EXAM}. 
% We compare their average \textit{Rank} and \textit{EXAM} in the 30 most popular SBFL ranking metrics.
The detailed ranking results of each case can be found on our website~\cite{website}.
% In this experiment, \tool used the \textit{arithmetic mean} function.

% \usepackage{multirow}
\begin{table*}[]
\scriptsize
\centering
\caption{Performance of \tool, SBFL, the combination of Slicing and SBFL (\textit{S-SBFL}), and Arrieta et al.~\cite{arrieta2018spectrum} (\textit{FB})}
\label{Table:comparison}
\begin{tabular}{|l|l|l|l|l|l|l|l|l|l|}
\hline
\multirow{2}{*}{\textbf{\#}} & \multirow{2}{*}{\textbf{Ranking Metric}} & \multicolumn{4}{c|}{\textit{\textbf{Rank}}}                              & \multicolumn{4}{c|}{\textit{\textbf{EXAM}}}                              \\ \cline{3-10} 
                             &                                  & \textbf{\tool} & \textbf{S-SBFL} & \textbf{SBFL} & \textbf{FB} & \textbf{\tool} & \textbf{S-SBFL} & \textbf{SBFL} & \textbf{FB} \\ \hline
\rowcolor[HTML]{C0C0C0}
1                            & Barinel                          & 7.83            & 9.88            & 11.48         & 136.27      & 2.11            & 2.87            & 3.15          & 21.79       \\ \hline
\rowcolor[HTML]{C0C0C0}
2                            & Dstar                            & 6.16            & 7.20            & 8.09          & 108.78      & 1.77            & 1.94            & 2.02          & 15.88       \\ \hline
\rowcolor[HTML]{C0C0C0}
3                            & Ochiai                           & 6.19            & 7.25            & 8.14          & 109.91      & 1.77            & 1.95            & 2.03          & 16.10       \\ \hline
\rowcolor[HTML]{C0C0C0}
4                            & Op2                              & 5.86            & 6.07            & 6.74          & 106.99      & 1.71            & 1.75            & 1.80          & 15.36       \\ \hline
\rowcolor[HTML]{C0C0C0}
5                            & Tarantula                        & 6.96            & 9.88            & 11.48         & 136.27      & 1.98            & 2.87            & 3.15          & 21.79       \\ \hline
6                            & Kulczynski2                      & 5.61            & 6.36            & 7.08          & 108.23      & 1.67            & 1.77            & 1.83          & 15.59       \\ \hline
7                            & M2                               & 5.94            & 6.13            & 6.82          & 108.43      & 1.71            & 1.76            & 1.81          & 15.77       \\ \hline
8                            & Harmonic Mean                    & 5.95            & 6.52            & 7.28          & 149.70      & 1.72            & 1.80            & 1.86          & 21.37       \\ \hline
9                            & Zoltar                           & 6.00            & 6.12            & 6.78          & 107.57      & 1.68            & 1.75            & 1.80          & 15.45       \\ \hline
10                           & Geometric Mean                   & 6.05            & 7.37            & 8.29          & 149.70      & 1.76            & 1.99            & 2.09          & 21.37       \\ \hline
11                           & Ample2                           & 6.15            & 6.16            & 6.86          & 149.58      & 1.75            & 1.77            & 1.82          & 21.30       \\ \hline
12                           & Rogot2                           & 6.22            & 6.52            & 7.28          & 133.66      & 1.80            & 1.80            & 1.86          & 22.24       \\ \hline
13                           & Sorensen Dice                    & 6.50            & 8.79            & 10.17         & 115.72      & 1.84            & 2.41            & 2.62          & 17.29       \\ \hline
14                           & Goodman                          & 6.50            & 8.79            & 10.17         & 115.72      & 1.84            & 2.41            & 2.62          & 17.29       \\ \hline
15                           & Jaccard                          & 6.63            & 8.79            & 10.17         & 115.72      & 1.83            & 2.41            & 2.62          & 17.29       \\ \hline
16                           & Dice                             & 6.63            & 8.79            & 10.17         & 115.72      & 1.83            & 2.41            & 2.62          & 17.29       \\ \hline
17                           & Anderberg                        & 6.68            & 8.79            & 10.17         & 115.72      & 1.84            & 2.41            & 2.62          & 17.29       \\ \hline
18                           & Cohen                            & 6.81            & 8.93            & 10.33         & 152.04      & 1.87            & 2.47            & 2.70          & 21.61       \\ \hline
19                           & Fleiss                           & 6.82            & 12.24           & 52.03         & 145.70      & 2.09            & 3.51            & 9.13          & 21.65       \\ \hline
20                           & Simple Matching                  & 6.88            & 28.00           & 242.70        & 158.19      & 2.11            & 6.67            & 30.68         & 21.96       \\ \hline
21                           & Humman                           & 6.88            & 28.00           & 242.70        & 158.19      & 2.11            & 6.67            & 30.68         & 21.96       \\ \hline
22                           & Wong2                            & 6.88            & 28.00           & 242.70        & 158.19      & 2.11            & 6.67            & 30.68         & 21.96       \\ \hline
23                           & Hamming                          & 6.88            & 28.00           & 242.70        & 158.19      & 2.11            & 6.67            & 30.68         & 21.96       \\ \hline
24                           & Sokal                            & 6.91            & 28.00           & 242.70        & 158.19      & 2.15            & 6.67            & 30.68         & 21.96       \\ \hline
25                           & Euclid                           & 6.96            & 28.00           & 242.70        & 158.19      & 2.17            & 6.67            & 30.68         & 21.96       \\ \hline
26                           & Rogers Tanimoto                  & 7.05            & 28.00           & 242.70        & 158.19      & 2.20            & 6.67            & 30.68         & 21.96       \\ \hline
27                           & Scott                            & 7.38            & 13.22           & 50.86         & 147.65      & 2.17            & 3.76            & 8.79          & 22.23       \\ \hline
28                           & Rogot1                           & 7.38            & 13.22           & 50.86         & 147.65      & 2.17            & 3.76            & 8.79          & 22.23       \\ \hline
29                           & Russell Rao                      & 14.06           & 17.87           & 24.00         & 309.62      & 3.58            & 5.05            & 6.39          & 39.27       \\ \hline
30                           & Wong1                            & 14.06           & 17.87           & 24.00         & 309.62      & 3.58            & 5.05            & 6.39          & 39.27       \\ \hline
\end{tabular}
\end{table*}

% \subsubsection{Performance comparison} 
\label{sec:performace_comparison}

\textbf{Compare to SBFL and S-SBFL.}
%
%Result
\textit{For both Rank and EXAM, \tool outperformed S-SBFL and SBFL in \textbf{all} the studied metrics}.
On average, \tool achieved \textbf{33\%} better in \textit{Rank} compared to S-SBFL and nearly \textbf{50\%} compared to SBFL. This means, to find a variability bug, by using \tool, developers have to investigate only 5 statements instead of about 8 and up to 10 suspicious statements by S-SBFL and SBFL.
In \textit{EXAM}, the improvements of \tool compared to both S-SBFL and SBFL are also significant, \textbf{30\%} and \textbf{43\%}, respectively. 
For developer, the proportion of statements they have to examine is reduced by about one third and one haft by using \tool compared to S-SBFL and SBFL.
For the 5 most popular metrics~\cite{pearson2017evaluating}, \cite{wen2019historical} (in shade) including Tarantula~\cite{taratula}, Ochiai~\cite{ochiai1957zoogeographic},  Op2~\cite{ding2013fault}, Dstar~\cite{wong2013dstar}, and Barinel~\cite{abreu2009spectrum}, 
% compared to both S-SBFL and SBFL, 
\tool achieved the improvement of more than \textbf{15\%}.
% in \textit{EXAM}.
%
Especially, for certain metrics such as Simple Matching~\cite{meyer2004comparison}, the improvements by \tool are remarkable, \textbf{4 times} and \textbf{35 times} compared to S-SBFL and SBFL.
%

%Example
\begin{figure}
    \centering
\begin{lstlisting}[language=Java]
public  boolean consistent(){
    for (int i = 0; i < students.length; i++) {
        if (students[++i] != null &&                     !students[i].backedOut &&                       students[i].points < 0) {
            <@\textcolor{ao}{//Patch: students[i] != null}@>
            return false;
        }
    }
    return true;
}
\end{lstlisting}
    \caption{A variability bug in system \textit{ExamDB}}
    \label{fig:example_comparison}
\end{figure}{}

%Explain
There are two reasons for these improvements.
\textit{Firstly, the set of the suspicious statements isolated by \tool is much smaller than other approaches'}. \tool identifies the suspicious statements by analyzing the root causes of failures. The suspicious space by \tool is only about 70\% of the space of S-SBFL and 10\% of the space of SBFL. The average suspicious space isolated by \tool is only 66 statements, while the isolated set by  S-SBFL contains 87 statements, and this suspicious set identified by SBFL is even much larger, 660 statements. 
\textit{Secondly, the suspiciousness of statements computed by \tool is not biased by the tests in any specific product.} 
Unlike SBFL, in \tool, 
for a suspicious statement $s$, the appearances of $s$ in both passing and failing products, as well as the local test case-based suspiciousness scores of $s$ in all the failing products containing it are aggregated appropriately.
% the contributions of a suspicious statement to the failures of the failing products containing it are locally measured in these products and globally aggregated appropriately. 
This suspiciousness measurement approach helps \tool overcome the weakness of SBFL in computing suspiciousness for the statements of systems.
%

%Example description
Fig.~\ref{fig:example_comparison} shows a variability bug (ID\_298) in feature \textit{BackOut} of \textit{ExamDB}. In this code, each member in \texttt{students} must be visited. However, the students with the even index are incorrectly ignored because \texttt{i} is increased twice after each iteration (line 2 and line 3). This bug is revealed only when both \textit{ExamDB} and \textit{BackOut} are enabled.
For QA, 8 products are sampled for testing. There are 4 failing products and total 168 statements executed during running failed tests. By using Tarantula~\cite{taratula}, \tool ranked the buggy statement  $s_3$ (line 3) first, while SBFL ranked it $10^{th}$.
%Analysis
Indeed by locally ranking the buggy statement in each product, $s_3$ is ranked $1^{st}$ in 3 out of 4 failing products. Meanwhile, this statement is ranked at the $27^{th}$ position in the ranked list of the other failing product ($p_4$).
%
% The reason is that the quality of the test suite of $p_4$ is quite bad. 
In $p_4$, there are 10 correct statements which are executed by failed tests only, yet not executed by any passed test. By using Tarantula, SBFL assigned these statements the highest suspiciousness scores in $p_4$. Thus, those statements have higher scores than the buggy statement which is executed during running both the failed and passed tests.
In the whole system, SBFL uses the test results of all the sampled products to measure the suspiciousness for all the 168 statements. Consequently, it misleadingly assigned higher scores for all of the 10
% 13 correct statements, and ranked the buggy one at $14^{th}$.
% Indeed, there are 10/13 
statements which are executed by only the failed tests in $p_4$, yet not executed by any tests in the others.
Consequently, the ranking result by SBFL is considerably driven by the test results of $p_4$ and mislocates the buggy statement. 
Meanwhile, \tool ignored 101/168 failure-unrelated statements and measured the suspiciousness of only 67 statements by analyzing the root cause of the failures. Additionally, in \tool, the test case-based suspiciousness of these statements are aggregated from the suspiciousness values which are measured in the failing products independently. Thus, the low-quality test suite of $p_4$ cannot significantly affect the suspiciousness measurement, and the buggy statement is still ranked $1^{st}$ thanks to the test suites of other products.

\begin{figure}[ht]
    \centering
    \includegraphics[width=0.95\linewidth]{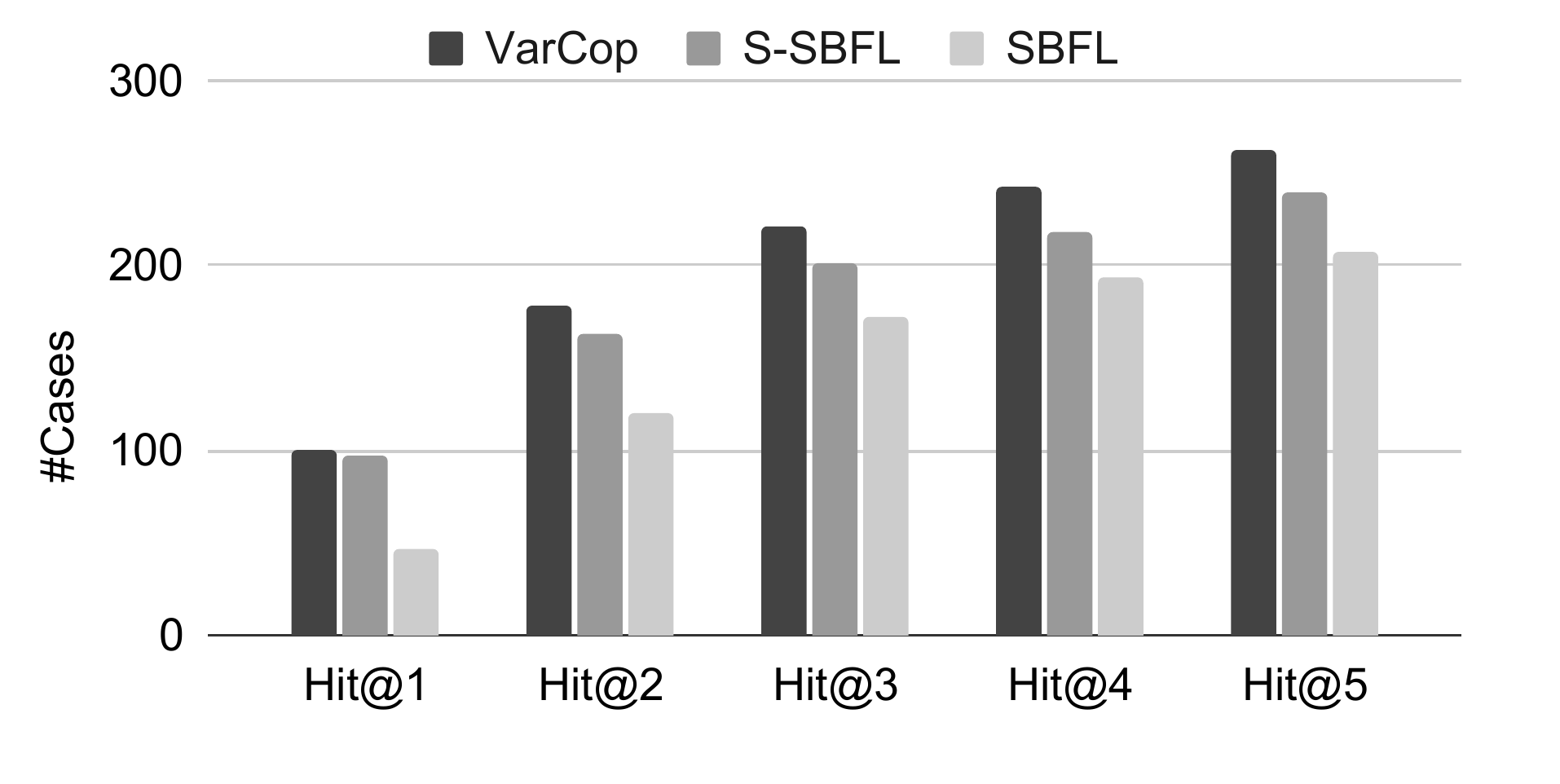}
    % \caption{The numbers of bugs found after examining 1--5 statements using \tool and SBFL, \textit{Hit@1--Hit@5}}
    \caption{\textit{Hit@1--Hit@5} of \tool, S-SBFL and SBFL}
    \label{fig:hit@x}
\end{figure}
%Hit@X
Furthermore, \textit{\tool also surpasses S-SBFL and SBFL in \textit{Hit@X}}.
%
% Fig.~\ref{fig:hit@x} shows the average \textit{Hit@X}, with $X \in [1, 5]$, of \tool and SBFL using 30 SBFL ranking metrics. 
%
In Fig.~\ref{fig:hit@x}, after investigating $X$ statements, \textit{for $X \in [1, 5]$, there are more bugs found by using \tool compared to S-SBFL and SBFL}. 
On average, in \textbf{78\%} of the cases, \tool correctly ranked the buggy statements at the top-5 positions, while S-SBFL and SBFL ranked them at the top-5 positions in only 70\% and 61\% of the cases, respectively. 
Moreover, in about two-thirds of the cases (\textbf{+65\%}), the bug can be found by examining \textbf{only first 3 statements} in the lists of \tool. Meanwhile, to cover the same proportion of the cases by using S-SBFL and SBFL, developers have to investigate up to 4 and 5 statements. 
%
%, which is \textit{double} SBFL's \textit{Hit@1}, about only 46 bugs
Especially, for \textit{Hit@1}, the number of bugs are found by \tool after investigating the first ranked statements is about \textbf{101 bugs (30\%)}. This means, in \textbf{one-third of the cases}, developers just need to examine the first statements in the ranked lists to find bugs by using \tool. 
%
% Particularly, Fig. \ref{fig:hit@1} shows \textit{Hit@1} of \tool and SBFL using 5 most popular metrics~\cite{pearson2017evaluating}, \cite{wen2019historical}. As seen, \tool outperforms SBFL in all of these metrics from \textbf{65--163\%} relatively.  

% Note that, among the subject systems, the average improvements of \tool compared to SBFL are not significantly different. For the studied systems, \tool's performance improvements are from 23\% to 55\% in \textit{Rank}.

% \begin{figure}
%     \centering
%     \includegraphics[width=0.85\linewidth]{Image/comparison/hit@1_v2.pdf}
%     \caption{\textit{Hit@1} of \tool and SBFL}
%     % \caption{\textit{Hit@1} of \tool and SBFL using 5 most popular SBFL ranking metrics}
%     \label{fig:hit@1}
% \end{figure}

\textbf{Compare to Arrieta et al.\cite{arrieta2018spectrum}.}
As illustrated in Table~\ref{Table:comparison},  in all the studied metrics, \tool outperformed Arrieta et al. \cite{arrieta2018spectrum} \textbf{21 times} in \textit{Rank} and \textbf{11 times} in \textit{EXAM}.
%
%Explain
% In fact, the average rank of the buggy features, which contain the buggy statements, is about \textbf{3.63}. 
Instead of ranking statements, this approach localizes the variability bugs at the feature-level. Consequently, all the statements in the same feature are assigned to the same score. In Fig.~\ref{fig:example_comparison}, the buggy statement at line 3 is assigned the same suspiciousness level with 22 correct statements. Thus, even the feature containing the fault, \textit{BackOut}, is ranked first, the buggy statement is still ranked $22^{nd}$ in the statement-level fault localizing. 
Unfortunately, \textit{BackOut} is actually ranked $4^{th}$, then the buggy statement is ranked $87^{th}$.
This could lead to the ineffectiveness of the feature-based approach proposed by Arrieta et al.\cite{arrieta2018spectrum} in localizing variability bugs in the statement-level.

\textit{Overall, our results show that {\tool} significantly outperformed the state-of-the-art approaches, S-SBFL, SBFL, and Arrieta et al.~\cite{arrieta2018spectrum}, in all 30/30 SBFL ranking metrics.}

\subsubsection{Performance by Bug Types}
%Varcop config: Op2, normalization1, aggregation4
We further analyzed \tool's performance on localizing bugs in different types based on mutation operators~\cite{ma2006mujava} and kinds of code elements~\cite{sobreira2018dissection}.
%
% In this experiments, \tool is configured with Op2~\cite{ding2013fault}, arithmetic mean aggregation function, and combination weight $w=0.5$.
%
%

In Table \ref{Table:bug_type_operators}, \tool performs most effectively on the bugs created by \textit{Conditional Operators} which are ranked between $1^{st}$ and $2^{nd}$ on average. 
%reason
%
The reason is that these bugs are easier to be detected (killed) by the original tests than other kinds of mutants~\cite{smith2009guiding}.
This means, if the bugs in this kind can cause the failures in some products, the bugs will be easier to be revealed by the products' tests. Moreover, the correct states of either passing or failing of products affect the performance of FL techniques.
% more details will be explained in Sec.~\ref{sec:sensitivity_test}.
As a result, this kind of bug is more effectively localized by \tool.
%
%\footnote{These operators delete a part of a statement.}
%
Meanwhile, \tool did not localize well the bugs created by \textit{Arithmetic Operators}, as they are more challenging to be detected by the original tests set of the products~\cite{smith2009guiding}. Indeed, because of the ineffectiveness of the test suites in several products, even they contain the bug(s), their test suites cannot detect the bugs, and the products still pass all their tests. In these cases, the performance would be negatively affected.
%~\cite{smith2009guidig}.   
%

\begin{table}[]
\centering
\caption{Performance by Mutation Operators}
\label{Table:bug_type_operators}
\begin{tabular}{|l|l|l|l|l|}
\hline
\textbf{Group} & \textbf{Mutation Operator}                                                             & \textbf{\#Bugs} & \textit{\textbf{Rank}} & \textit{\textbf{EXAM}} \\ \hline

Conditional    & \textit{COR, COI, COD}                                                                 & 32              & 1.63                   & 0.61\\ \hline
Assignment     & \textit{ASRS}                                                                          & 7              & 2.14                   & 0.89                   \\ \hline

Logical        & \textit{LOI}                                                                           & 17              & 2.47                   & 2.21                   \\ \hline
Deletion       & \textit{CDL, ODL}                                                                      & 18              & 3.56                   & 1.09                   \\ \hline

Relational     & \textit{ROR}                                                                           & 52              & 5.13                  & 1.29                   \\ \hline

Arithmetic     & \textit{\begin{tabular}[c]{@{}l@{}}AODU, AOIU, AORB, \\ AOIS, AORS, AODS\end{tabular}} & 212              & 7.63                  & 2.01                   \\ \hline

\end{tabular}
\end{table}

Table \ref{Table:bug_type_code_elements} shows  \tool's performance in different kinds of bugs categorized based on code elements~\cite{sobreira2018dissection}. As seen, \textit{\tool works quite stably in these kinds of bugs}. Particularly, the average \textit{Rank} achieved by \tool for bugs in different code elements is between $4^{th}$ and $7^{th}$, with the standard deviation is only 1.3. In addition, the average $EXAM$ and the standard deviation are 1.53 and 0.73, respectively.

% \subsubsection{Performance by Bugs' Code Elements}

% However, the average \textit{Rank} in \textit{Return} statements is worse than the others'. There is only 1 out of 35 cases in this kind where \tool did not work well. The \textit{Rank} for this case is \textbf{93}, while the average figure of the remaining cases in \textit{Return} category is \textbf{3.68}. 
% %
% This low-performance case is a buggy version of \textit{GPL} system.
% For this case, in method \texttt{getWeight}, the \textit{Return} statement returns the expression with a decreasing postfix operator, \texttt{this.Weight--}. Thus, \texttt{getWeight} unexpectedly decreased \texttt{Weight} \textit{after returning its value}. This bug is very tricky to be found by all FL techniques because its visibility depends on whether \texttt{Weight} is used after \texttt{getWeight} is invoked or not. If \texttt{getWeight} is invoked once, the returned value of this method is still correct. 
% However, when \texttt{getWeight} is invoked more than once, \texttt{Weight} is incorrectly decreased after the first call, and \texttt{getWeight} returns an unexpected value. 
% For this case, 
% SBFL ranked the bug, even more ineffectively, at the 98th position in its list.

\begin{table}[]
\centering
\caption{Performance by Code Elements of Bugs}
\label{Table:bug_type_code_elements}
\begin{tabular}{|l|l|l|l|}
\hline
\textbf{Code Element}            & \textbf{\#Bugs} & \textbf{\textit{Rank}} & \textbf{\textit{EXAM}} \\ \hline
Method Call &  22    & 4.23 & 0.37\\ \hline
Conditional & 148     & 5.20 & 1.86 \\ \hline
Loop        & 17     & 6.41 & 2.27 \\ \hline
Assignment  & 108     & 7.18 & 1.82 \\ \hline
Return      & 43     & 7.21 & 1.33 \\ \hline

\end{tabular}
\end{table}

\subsubsection{Performance by the number of involving features}

\begin{figure}
    \centering
    \includegraphics[width=0.9\linewidth]{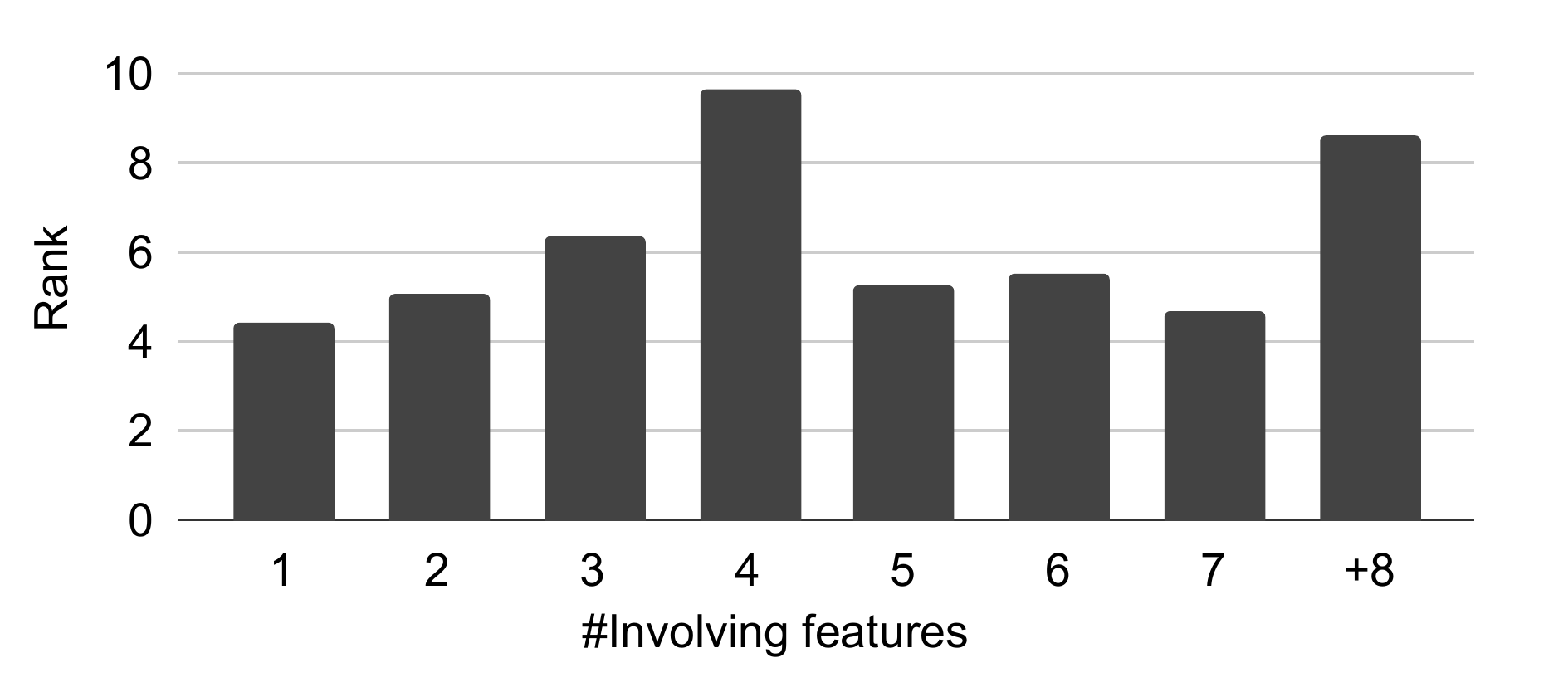}
    \caption{Performance by number of involving features of bugs}
    \label{fig:involving_features}
\end{figure}

We also analyzed the performance of \tool by the number of the features which involve in the visibility of the bugs~\cite{ourdataset}.
% must be \textit{actually} enabled/disabled to reveal bugs (involving features). 
% %
% A feature is an involving feature to a bug if from a failing product when switching its current selection makes the resulting product pass all its tests. If the resulting product has not been sampled, we additionally compose the product and generate its tests.
% %
%
In our experiment, the number of involving features is in the range of $[1, 25]$. In +76\% of the cases, the number of involving features is fewer than or equal to 7. \tool's performance in \textit{Rank} by the numbers of involving features are not significantly different, from 4.45 to 9.69, Fig.~\ref{fig:involving_features}.
In fact, both the number of detected \spcs and the size of each \spc, which determine the size of isolated suspicious space, are affected by the number of involving features.
Specially, for the bugs with a smaller number of involving features, the detected \spcs are likely fewer, but each of them is likely smaller. With a larger number of involving features, the detected \spcs tend to be more, yet each of these \spcs is likely larger. 
For a bug, the isolated suspicious statements space is in direct proportion to the number of detected \spcs, but it is in inverse proportion to the size of each \spc. 
Thus, the number of involving features would not linearly affect the number of isolated suspicious statements and \tool's performance. 
%
% For detailed results, see our website~\cite{website}. 
%
% Thus, for a bug, the isolated suspicious statements space is in direct proportion to the number of detected \spcs, but it is in inverse proportion to the size of each \spc. 

\subsection{Intrinsic Analysis (RQ2)}
\label{sec:intrisic}

\subsubsection{Impact of Suspicious Statements Isolation on Performance}
\label{section:impact_bpc_detection}

\begin{figure}
    \centering
    \includegraphics[width=0.95\linewidth]{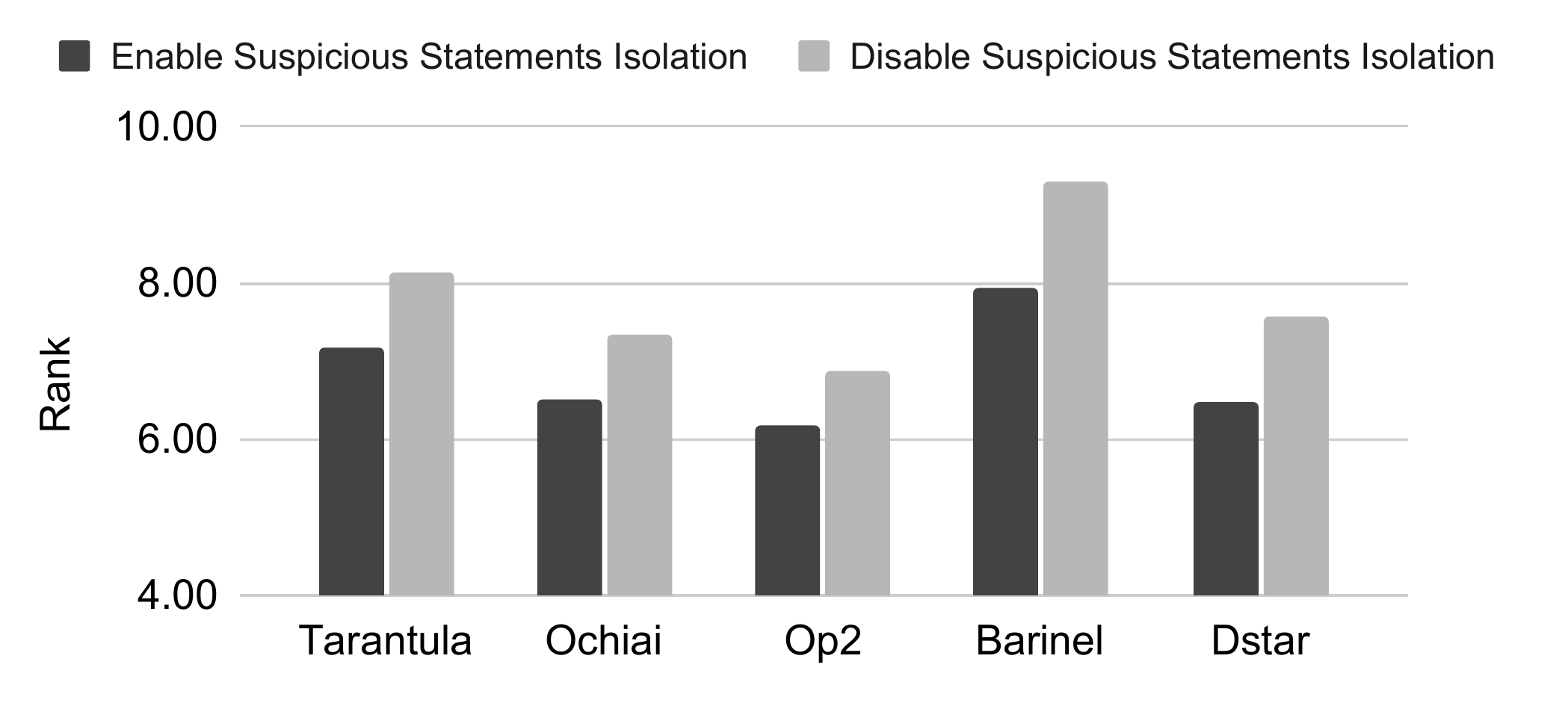}
    \caption{Impact of \bpc Detection on performance}
    \label{fig:intrinsic_bpc}
\end{figure}

To study the impact of Suspicious Statements Isolation (Fig.~\ref{fig:varcop_overview}), which includes \bpc Detection and Suspicious Statements Identification components, on \tool's performance, we built the variant of \tool where these two components are disabled. For a buggy system, this variant of \tool ranks all the statements which are executed during running failed tests in the failing products.
% without \bpc Detection and Suspicious Statement Isolation.
%
Fig.~\ref{fig:intrinsic_bpc} shows the performance of \tool using 5 most popular SBFL metrics~\cite{pearson2017evaluating}, \cite{wen2019historical} when \bpc Detection and Suspicious Statements Identification are enabled/disabled. As expected, \textit{when enabling these components, the performance of \tool is significantly better, about 16\% in \textit{Rank}}.

% Especially, when enabling BPC Detection, \tool using Wong3~\cite{?} obtains \textbf{about 2.5 times better} on average in \textit{ranking} compared to \tool when this component is disabled. 

% \begin{figure}[ht]
%     \centering
%     \begin{subfigure}{0.5\textwidth}
%     \centering
%         \includegraphics[width = 0.7\textwidth]{Image/intrinsic/intrinsic_bpc_rank.pdf}
%         %  \subcaption{Performance in \textit{Rank}}
%         \label{fig:intrinsic_bpc_rank}
%     \end{subfigure}\hfil 
%     % \begin{subfigure}{0.5\textwidth}
%     % \centering
%     %     \includegraphics[width = 0.9\textwidth]{Image/intrinsic/intrinsic_bpc_exam.pdf}
%     %     \subcaption{Performance in \textit{EXAM}}
%     %     \label{fig:intrinsic_bpc_exam}
%     % \end{subfigure}\hfil 
    
%       \caption{Impact of \bpc Detection on performance}
%       \label{fig:intrinsic_bpc}
% \end{figure}{}

Interestingly, \textit{even when disabling Suspicious Statements Isolation, this variant of \tool is still better than S-SBFL and SBFL}.
Specially, \tool obtained a better \textit{Rank} than S-SBFL and SBFL in 21/30 and 27/30 metrics, respectively. In these ranking metrics, the average improvements of \tool compared to S-SBFL and SBFL are 34\% and 45\%. 
Meanwhile, for the remaining metrics, the performances of S-SBFL and SBFL are better than \tool by only 10\% and 3\%.
%
% \footnote{The details can be found on our website.}.
%
% and using Tarantula\cite{taratula}, 
% \tool measures local suspicioueness for all the statements executed by failing tests (i.e. 168 statements) of system \textit{EXAMDB}. Then, global scores are calculated by maintaining the relative contribution of these statements to each failures, \tool is not significantly affected by test results any specific product. 
% Therefore, 
For the bug in Fig.~\ref{fig:example_comparison}, thanks to the proposed suspiciousness measurement approach, this variant of \tool still ranked the buggy statement ($s_3$) $1^{st}$ which is much better than the $9^{th}$ and $10^{th}$ positions by S-SBFL and SBFL.

Note that, for 19/338 cases which possibly contain non-variability bugs (mentioned in Section~\ref{sec:data}), there might be no \bpc in these buggy systems to be detected. Moreover, the low-quality test suites in some passing (yet buggy) products might ``fool" fault localization techniques~\cite{baudry2006improving}. These passing products might also make \tool less effective in isolating suspicious statements.  
% \tool and especially its suspicious statements isolation are designed for localizing variability bugs. 
% In addition, the suspicious statements are isolated based on both the passing and failing products. 
Hence, for these cases, we turn off \tool's suspicious statements isolation component to guarantee its effectiveness. 
% For instance, in this experiment, \tool turns off this component in 19/338 cases which are mentioned in Section~\ref{sec:data}.

\subsubsection{Impact of Ranking Metric on Performance}
% \textbf{2. Impact of Choosing Local Measurement Metric on Performance.}
We also studied the impact of the selection of the local ranking metric on \tool's performance. 
To do that, we built different variants of \tool with different metrics. 
% The performances of these variants are also shown in Table  \ref{Table:comparison}.
%
%
In Table~\ref{Table:comparison} (the $3^{rd}$ and $7^{th}$ columns), \textit{the performance of \tool is quite stable with the different ranking metrics}.
%
% Particularly, for all the ranking metrics, the average \textit{Rank} of the buggy statements assigned by \tool is in a narrow range, from \textbf{5th--9th} with a standard deviation of \textbf{1.5}.
Particularly, for all the studied metrics, the average \textit{EXAM} achieved by \tool is in a narrow range, from 1.71--3.58, with the standard deviation of 0.46. Additionally, the average \textit{Rank} of the buggy statements assigned by \tool is varied from $6^{th}$--$14^{th}$.
This stability of \tool is obtained due to the suspicious statements isolation and suspiciousness measurement components. Indeed, \tool only considers the statements that are related to the interactions which are the root causes of the failures. Moreover, \tool is not biased by the test suites of any specific products. Consequently, its performance is less affected by the low-quality test suites of any product.  
Thus, \textit{selecting an inappropriate ranking metric, which is unknown beforehand in practice, does not significantly affect \tool's performance}. This demonstrates that \tool is practical in localizing variability bugs.

In contrast, the performances of S-SBFL and SBFL techniques are considerably impacted by choosing the ranking metrics. By S-SBFL method, the average \textit{Rank} of the buggy statements widely fluctuates from $6^{th}$ to $28^{th}$.
% and the range of \textit{EXAM} is $[1.75, 6.67]$.
By SBFL, the fluctuation of \textit{Rank} is even much more considerable, from $7^{th}$ to $243^{rd}$.  
Consequently, the QA process would be extremely inefficient if developers using the SBFL technique with an inappropriate ranking metric.

\subsubsection{Impact of Normalization on Performance}
% \textbf{3. Impact of Normalization on Performance. }
%
To study the impact of the normalization, we built the variants of \tool which enable and disable the normalization component. In this experiment, in both cases, the local test case-based scores are accordingly measured by 30 popular SBFL metrics and are aggregated by \textit{arithmetic mean}.
%
% \footnote{Normalization functions clearly impact the performance of \tool when the local scores of the statements in the different products are significantly different}. 
%

% \begin{figure}[ht]
%     \centering
%     \begin{subfigure}{0.5\textwidth}
%     \centering
%         \includegraphics[width = 0.9\textwidth]{Image/intrinsic/intrinsic_normalization_rank.pdf}
%          \subcaption{Performance in \textit{Rank}}
%         \label{fig:intrinsic_normalization_rank}
%     \end{subfigure}\hfil 
%     \begin{subfigure}{0.5\textwidth}
%     \centering
%         \includegraphics[width = 0.9\textwidth]{Image/intrinsic/intrinsic_normalization_exam.pdf}
%         \subcaption{Performance in \textit{EXAM}}
%         \label{fig:intrinsic_normalization_exam}
%     \end{subfigure}\hfil 
    
%       \caption{Impact of Normalization on performance}
%      \label{fig:intrinsic_normalization}
% \end{figure}{}

\begin{figure}
    \centering
    \includegraphics[width=1\linewidth]{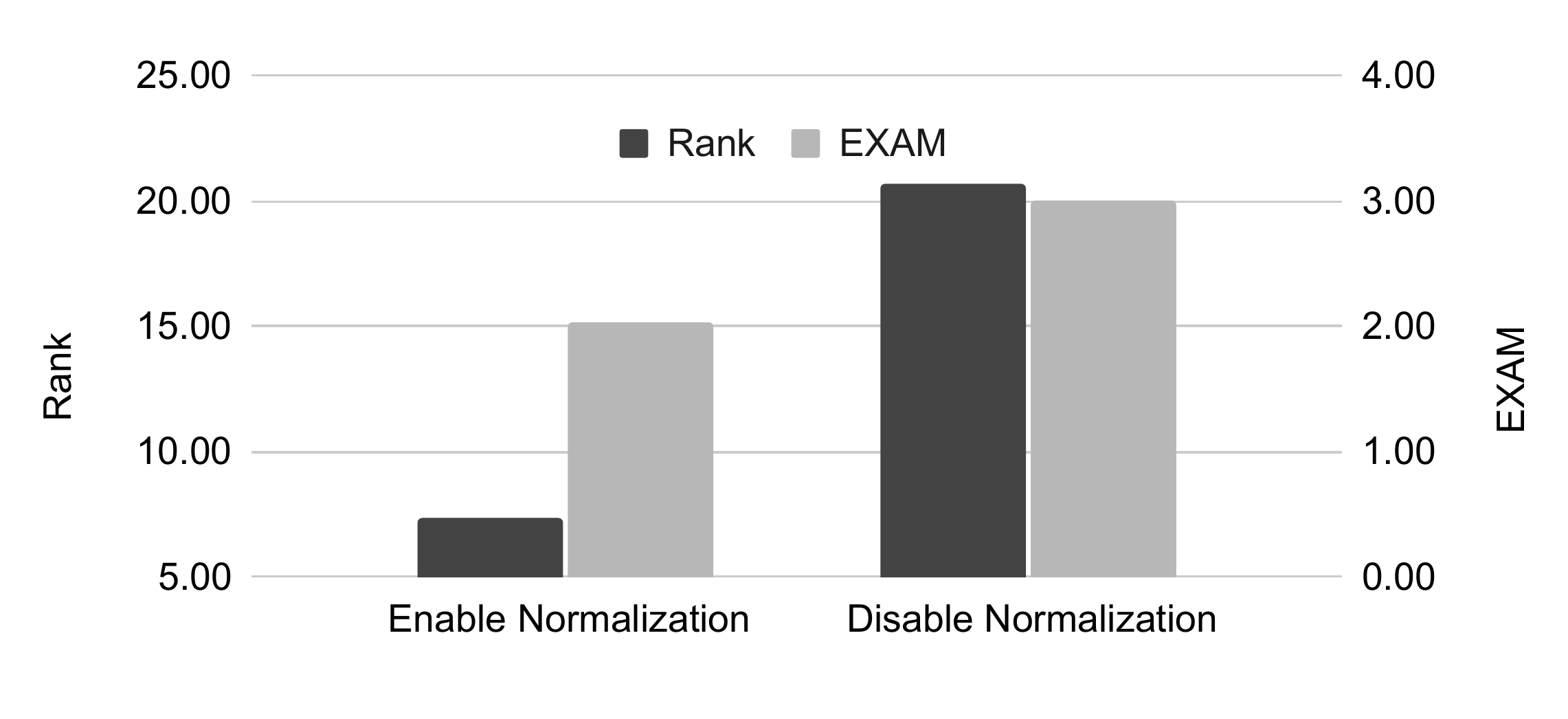}
    \caption{Impact of Normalization on performance}
     \label{fig:intrinsic_normalization}
\end{figure}

In Fig.~\ref{fig:intrinsic_normalization}, when enabling the \textit{normalization}, \tool's performance is better than when \textit{normalization} is off. Particularly, the performance of \tool is improved 64\% in \textit{Rank} and 32\% in \textit{EXAM} when the \textit{normalization} is enabled.
%and 21\% in \textit{EXAM}. 
%
One reason is that for some SBFL metrics, such as Fleiss~\cite{fleiss1965estimating}  and Humman~\cite{lourencco2004binary}, the ranges of the product-based and test case-based suspiciousness values are \textit{significantly different}. Additionally, for these metrics, the ranges of the local test case-based suspiciousness scores in different products are also \textit{significantly different}. For example, there is a bug (ID\_25) in the system \textit{Email}, with Fleiss, the range of suspiciousness scores in  product $p_1$ is $[-33, 1.8]$, while the range in another product, $p_2$ is much different, $[-8.6, 1.78]$. 
% After aggregating the suspiciousness scores according to test-cased based assessment and product-based assessment are in ranges $[-16,0.93]$ and $[-13.5, -0.52]$, respectively.
% %
Without \textit{normalization}, a statement in $p_2$ is more likely to be assigned a higher final score than one in $p_1$.
Meanwhile, with several metrics such as Ochiai~\cite{ochiai1957zoogeographic} and Tarantula~\cite{taratula}, the performance of \tool is slightly different when \textit{normalization} is on/off. %
For these metrics, the local scores of the statements in the products are originally assigned in quite similar ranges. Thus, these local scores might not need to be additionally normalized.   
\textit{Overall, to ensure that the best performance of \tool, the \textit{normalization} should be on}.
% to eliminate the potential inaccurate assessment caused by the difference in the ranges of the local suspiciousness scores.

\subsubsection{Impact of Aggregation Function on Performance}
\begin{figure}
    \centering
    \includegraphics[width=0.95\linewidth]{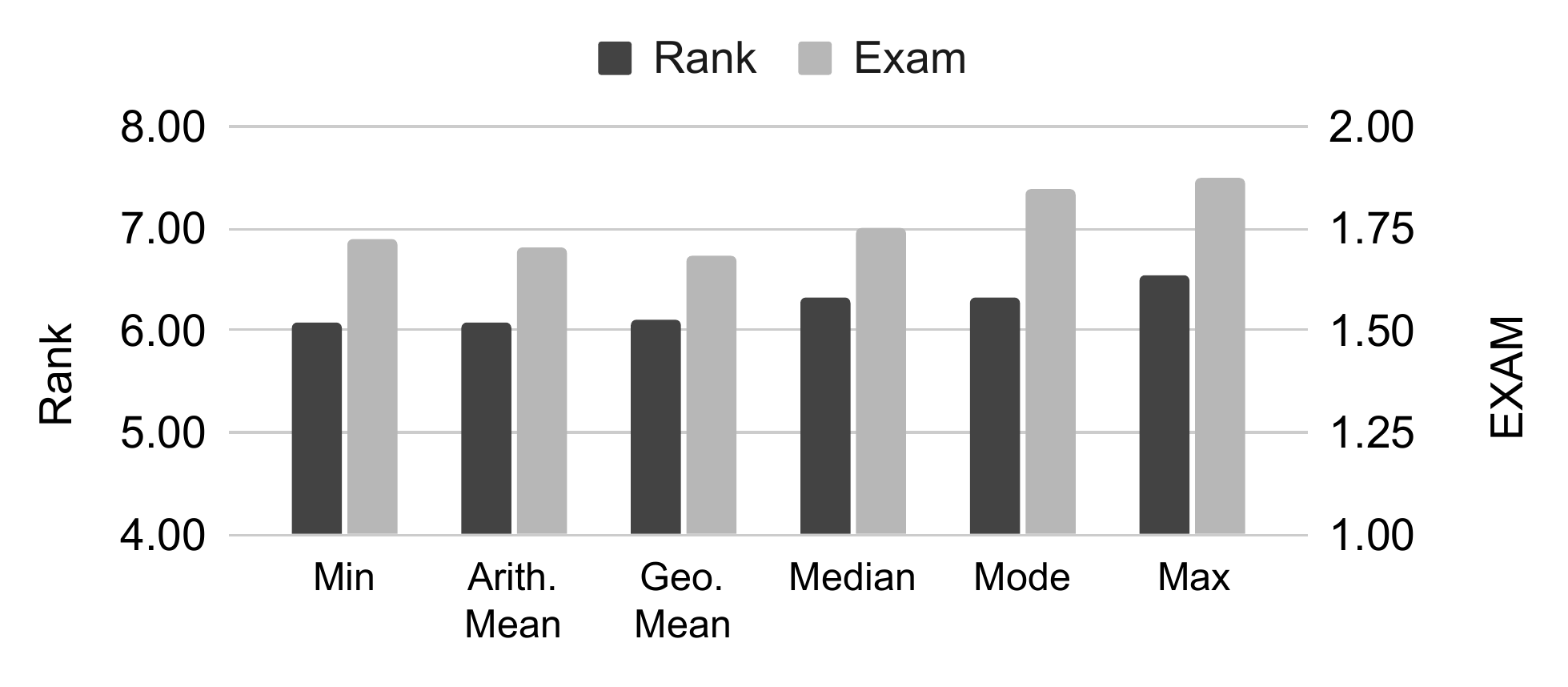}
     \caption{Impact of choosing $score(s, M)$ on performance}
      \label{fig:intrinsic_aggregation_functions}
\end{figure}

To study the impact of choosing aggregation function on performance, we varied the aggregation function of the test case-based suspiciousness assessment. In this experiment, Op2~\cite{ding2013fault} was randomly chosen to measure the local scores of the statements.
% After that, the normalized local scores are aggregated by different aggregation functions.
% as mentioned in Sec.~\ref{sec:aggregation_function}.  
%
As seen in Fig.~\ref{fig:intrinsic_aggregation_functions}, \textit{the performance of \tool is not significantly affected when the aggregation function is changed}. Specially, the average \textit{Rank} of \tool is around $6^{th} - 7^{th}$, while the \textit{EXAM} is about 1.76. 
%
% However, the performance of \tool using \textit{Max} is worse than using others.
% %
% This is because once a correct statement $s$ is assigned the highest score in a product due to its low-quality test suite, its normalized score in that product, which is in the common range of $[m_1, m_2]$, will be $m_2$. After being aggregated by \textit{Max}, $s$ will still have the highest global score, $m_2$, in the whole system. Consequently, even the buggy statement is correctly measured in other products, it will still be ranked below or equal to $s$. In the worst case, developers still examine the buggy statement after investigating $s$. Hence, \textit{Max} should not be used in \tool.

\subsubsection{Impact of Combination Weight on Performance}

\begin{figure}
    \centering
    \includegraphics[width=0.9\linewidth]{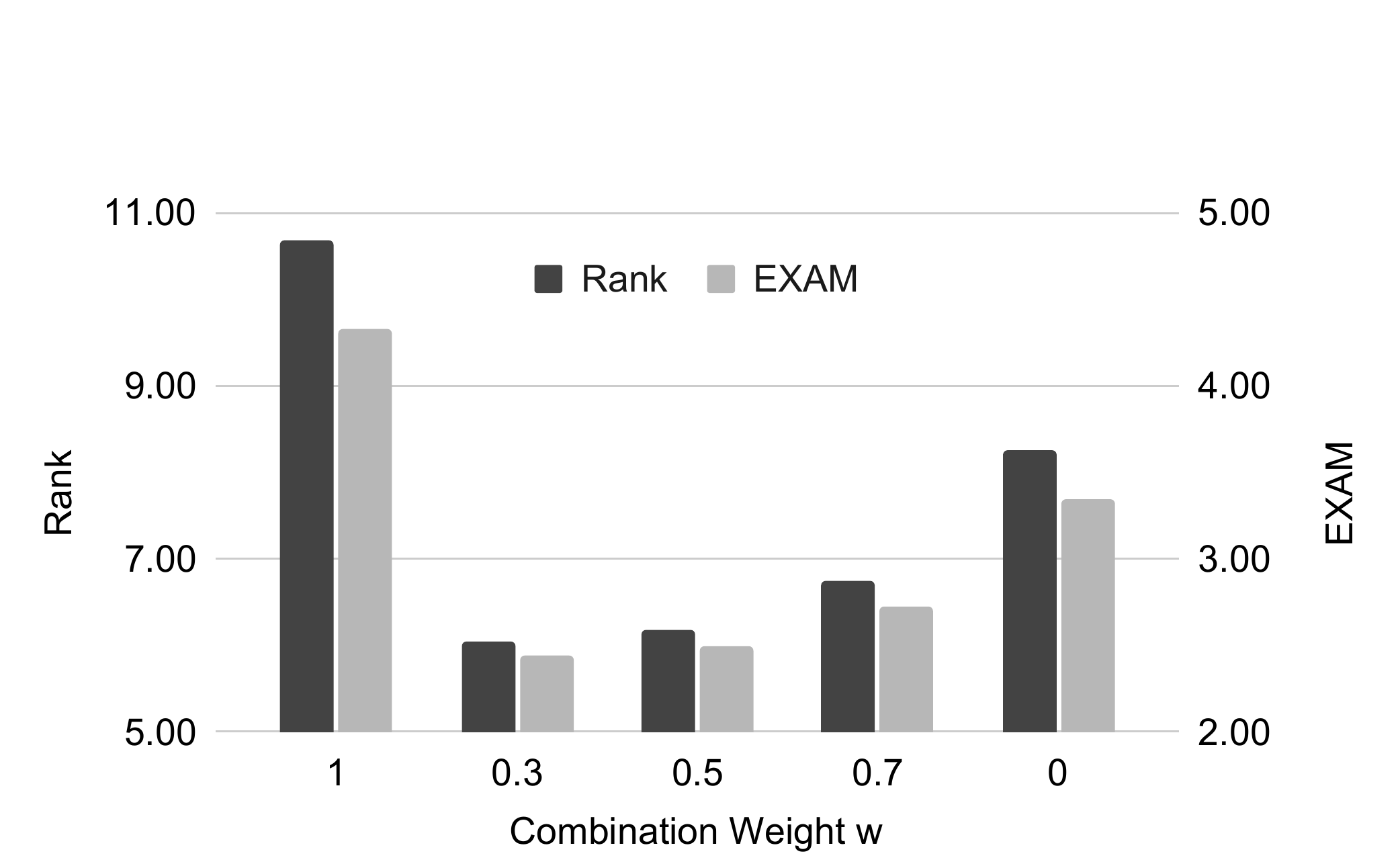}
     \caption{Impact of choosing combination weight on performance}
      \label{fig:intrinsic_combination_weight}
\end{figure}

We varied the \textit{combination weight} $w \in [0, 1]$ (Section \ref{sec:ranking}) when combining the product-based and test case-based suspiciousness assessments to form the final score of statements. Fig.~\ref{fig:intrinsic_combination_weight} shows the average \textit{Rank} and \textit{EXAM} of the faults in 36 buggy versions of \textit{Email}, with $w \in [0, 1]$. 
% Note that \tool is configured with Dstar and Arith. Mean.

As seen, \textit{the performance of \tool is better when both the product-based and test case-based suspiciousness scores are combined to measure the suspiciousness of the statements.} 
For $w=1$, the statements are ranked by only their product-based suspiciousness. All the statements in the same feature will have the same suspiciousness score because they appear in the same number of passing and failing products. 
For $w=0$, a statement is ranked by only the score which is aggregated from the local scores of the statement in the failing products. Consequently, the overall performance may be affected by the low-quality test suites of some products. For instance, the correct statement $s$ appears in only one failing product $p$. However, in $p$, $s$ is misleadingly assigned the highest score. As a result, when $w=0$, $s$ also has the highest score in the whole system, since this score is aggregated from $p$, the only failing product containing $s$. 
Hence, \textit{both of the product-based and test case-based suspiciousness assessments are necessary for measuring the suspiciousness of statements} (Mentioned in \textbf{O3}).

%- test coverage threshold
\subsection{Sensitivity Analysis (RQ3)}

\subsubsection{Impact of Sample Size on Performance}
% \noindent\textbf{1. Impact of Sample Size on Performance}

In this experiment, for each buggy version of \textit{GPL} which is randomly selected system, we used $k$-wise coverage, for $k \in [1,4]$, to systematically vary sample size. Then, we ran \tool on each case with each set of sampled products. 

\begin{figure}
    \centering
    \includegraphics[width=1\linewidth]{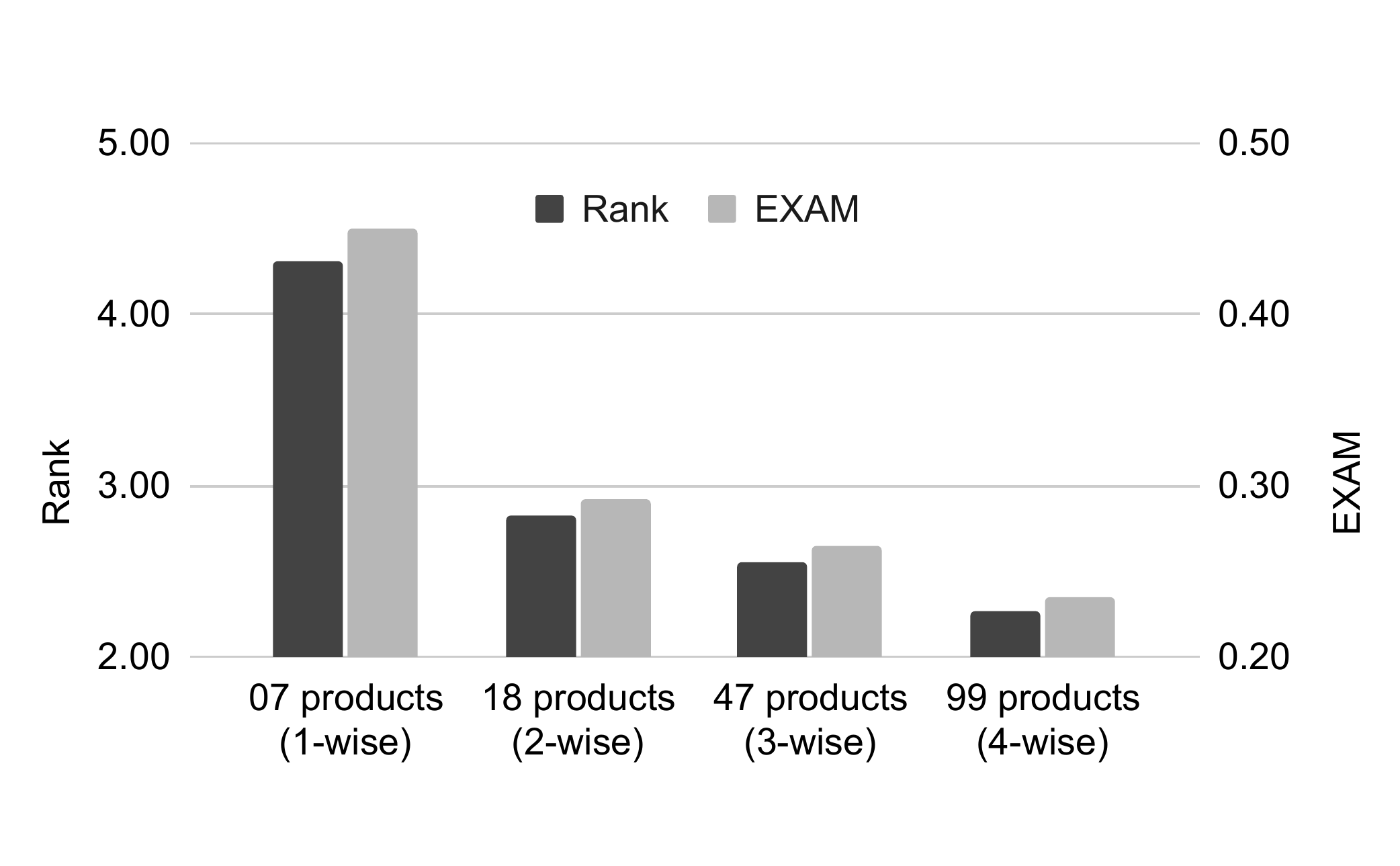}
    \caption{Impact of the sample size on performance}
    \label{fig:sensitivity_products}
\end{figure}

Fig.~\ref{fig:sensitivity_products} shows the average \textit{Rank} and \textit{EXAM} of \tool in the buggy versions of \textit{GPL} with different sample sets.
As expected, \textit{the larger sample, the higher performance in localizing bugs obtained by \tool}. However, when the ranking results reach a specific point, even more products are tested, the results are just slightly better. Specially, for the set of \textit{$1$-wise} coverage (\textit{One-disabled}~\cite{42bugs}), the average \textit{Rank} and \textit{EXAM} are about 4.31 and 0.45, respectively. Meanwhile, for \textit{$2$-wise}, the ranking results are 1.5 times better.
The reason is that, for a case, the more products are tested, the more information \tool has to detect \bpcs and rank suspicious statements. 
However, compared to \textit{3-wise} and \textit{4-wise}, even much more products are sampled, which is much more costly in sampling and testing, the performance is just slightly improved. 
Hence, \textit{with \tool, one might not need to use a very large sample to achieve a relatively good performance in localizing variability bugs.}

\subsubsection{Impact of Test Suite's Size on Performance}
\label{sec:sensitivity_test}

In this experiment, for every buggy version, we gradually increased the size of the test suite in each product to study the impact of tests on \tool's performance. The randomly selected system in this experiment is \textit{ExamDB}. 
% For every buggy version, in each product, the size of the test suite is gradually increased. 
%  which has 51 buggy versions
%

\begin{figure}
    \centering
    \includegraphics[width=0.93\linewidth]{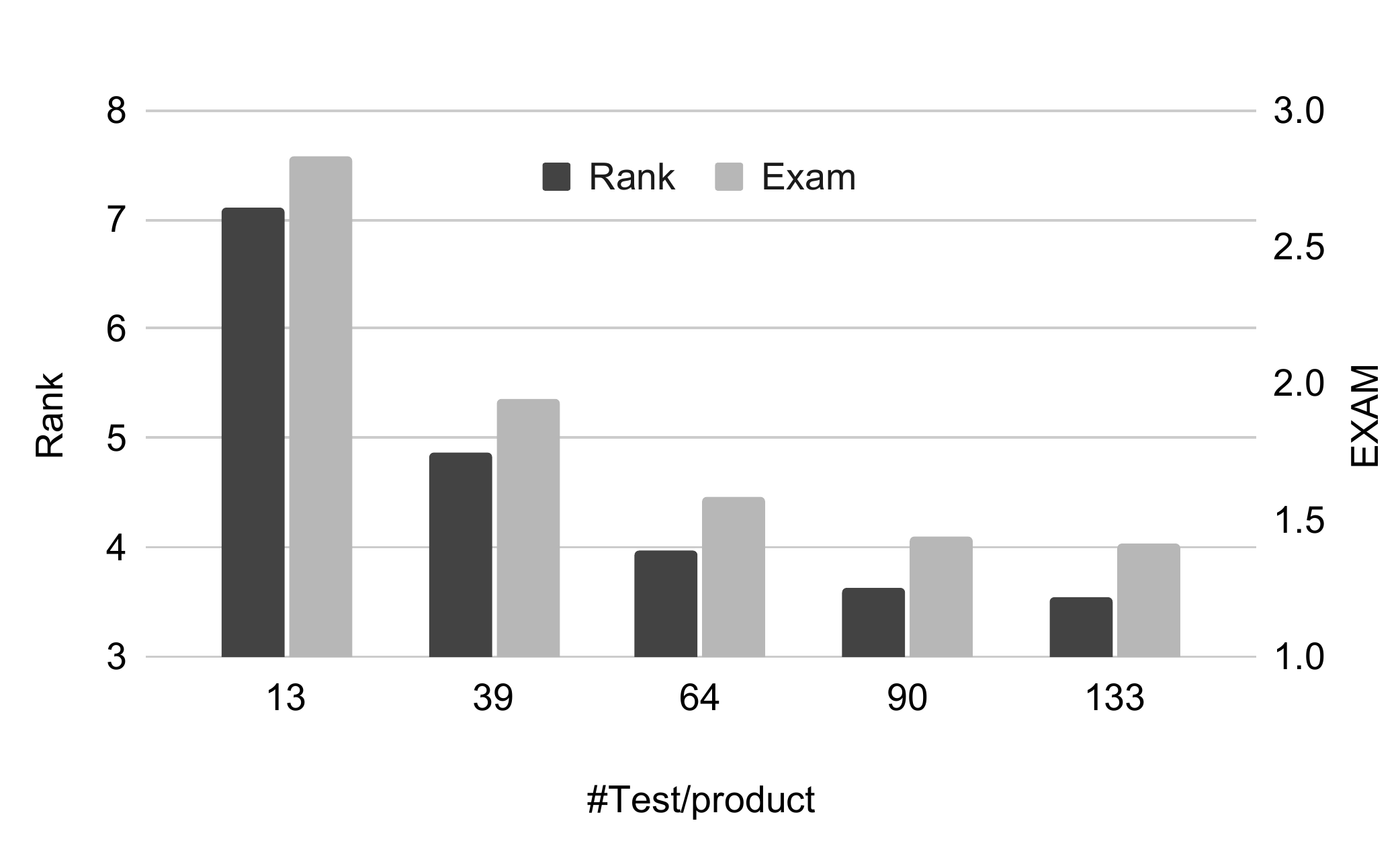}
    \caption{Impact of the size of test set on performance}
    \label{fig:sensitivity_tests}
\end{figure}

In Fig.~\ref{fig:sensitivity_tests}, \textit{\tool's performance is improved when increasing the test suite size.} Particularly, when the number of tests increased from 13 to 90 tests/product, both \textit{Rank} and \textit{EXAM} of \tool are improved by about twice. After that, even more tests are added, \tool's performance is just slightly affected. The reason is, increasing the number of tests provides more information to distinct the correct and incorrect statements, thus improves \tool's performance. However, when the test suites reach a specific effectiveness degree in detecting bugs, added tests would not provide supplementary information for the FL process.

% Moreover, we also found 1/51 case when adding more tests, \tool's performance is significantly better. In this case, originally, the test suite of a product is ineffective in detecting the bug. For instance, because a buggy product passes all the tests (i.e., a passing product), this product is misleadingly considered bug-free. With that information, \tool did not detect well \bpcs. 
% %
% Consequently, the bug is not ranked well because it's rank is lower than any other statements in the isolated suspicious statements set.
% With more tests, the test suites become more effective. Then, the incorrect statement is effectively captured by \tool.

Overall,\textit{ one should trade off between the fault localization effectiveness and the cost of generating tests and running them. Furthermore, as discussed in Section \ref{section:impact_bpc_detection}, instead of focusing on expanding test suites for products, developers should improve the effectiveness of the test suites in detecting bugs. This can be done by improving test coverage and applying mutation testing~\cite{mutation_testing}.}
\subsection{Performance in Localizing Multiple Bugs (RQ4)}
To evaluate \tool on buggy systems that contain multiple variability bugs, we conducted an experiment on 1,232 buggy versions of the subject systems with 2,947 variability bugs in total. 
% Each buggy version contains from 2 to 5 bugs.
%
The components of \tool were randomly configured: the ranking metric is Op2~\cite{ding2013fault}, and the aggregation function is \textit{arithmetic mean}. For other settings of \tool, one can find on our website~\cite{website}.

\begin{figure}
    \centering
    \includegraphics[width=1\linewidth]{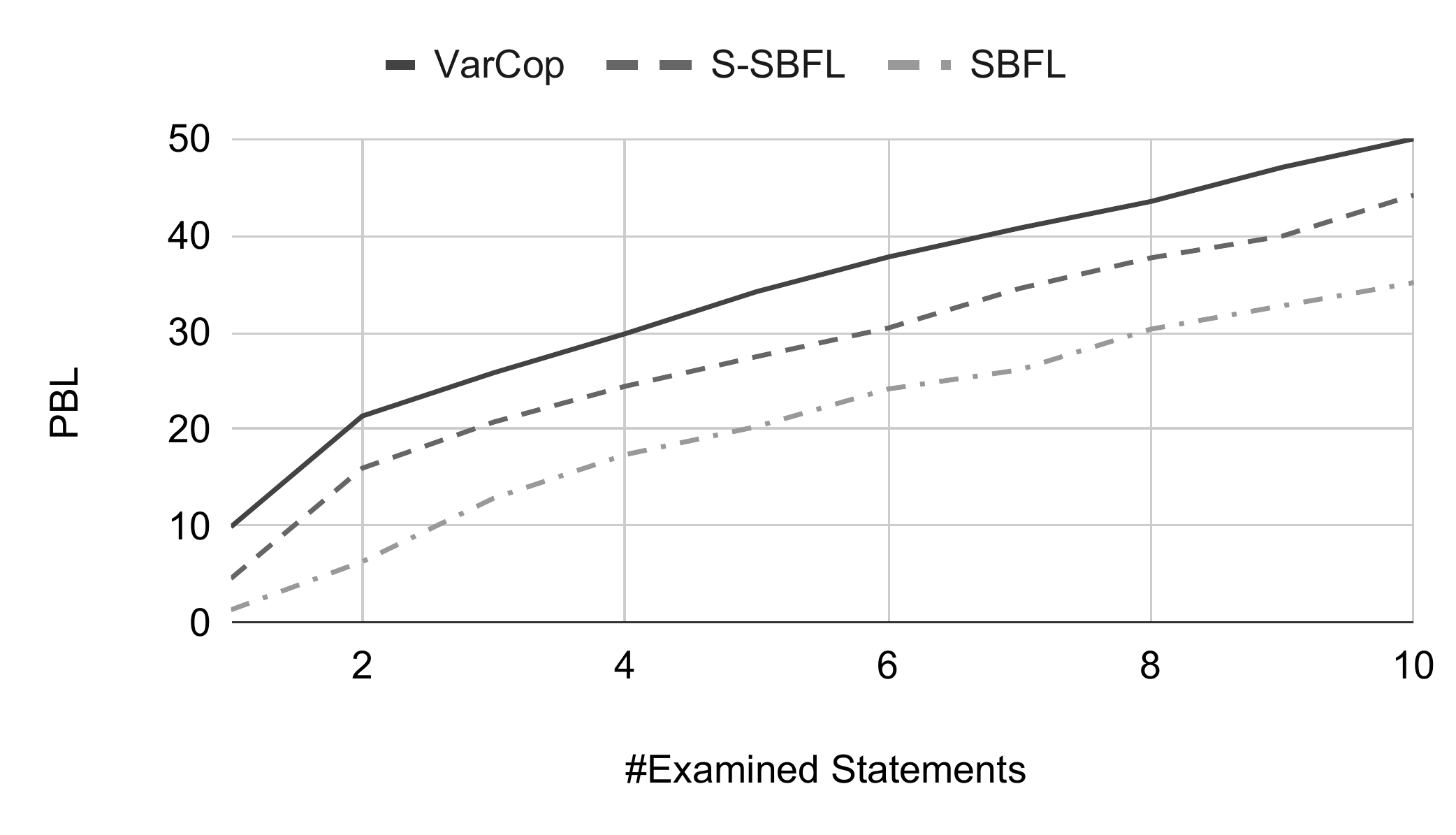}
    \caption{\tool, S-SBFL and SBFL in localizing multiple bugs}
    \label{fig:mb_percentage_of_bugs_found}
\end{figure}

Fig.~\ref{fig:mb_percentage_of_bugs_found} shows that the average percentage of buggy statements in each case found (PBL) by \tool far surpasses the corresponding figures of S-SBFL and SBFL when the same number of statements are examined in their ranked lists. Specially, after examining the first statement, \tool can find nearly \textbf{10\%} of the incorrect statements in a buggy system. Meanwhile, only 5\% and 1\% of the bugs are found by S-SBFL and SBFL, respectively, after inspecting the first statement. Furthermore, about \textbf{35\%} of the bugs can be found by \tool by checking only first \textbf{5} statements in the ranked lists. Meanwhile, with S-SBFL and SBFL, developers have to investigate up to 7 and even 10 statements to achieve the same performance.
In addition, the average \textit{best} \textit{Rank}
that \tool assigned for the buggy statements is about \textbf{$7^{th}$}. Meanwhile, the corresponding figures of S-SBFL, SBFL, and  Arrieta et al.~\cite{arrieta2018spectrum} are $7.5^{th}$, $10^{th}$, and $293^{th}$, respectively.

% Moreover, since in practice, the number of bugs in a system is unknown beforehand, developers have to conduct the testing and debugging processes in regression. In other words, they often fix and re-test the system right after a bug is found.
%
Especially, in \textbf{22\%} of the cases, \tool correctly ranked at least one of the bugs at the \textbf{top-1 positions}, while for S-SBFL, SBFL, and Arrieta et al.~\cite{arrieta2018spectrum}, the corresponding proportions are only 10\%, 3\%, and 0\%. In our experiment, at least one of the buggy statements of about \textbf{65\%} of cases are correctly ranked at top-5 positions by \tool.   

Fig.~\ref{fig:multiple_bugs_example} shows a case (ID\_143) containing two buggy statements in \textit{GPL}.
For the bug in Fig.~\ref{fig:multiple_bugs_example} (\subref{fig:mb_first}), \tool correctly ranked it $1^{st}$. 
% (at this time, the bug in Fig.~\ref{fig:multiple_bugs_example} (\subref{fig:mb_second}) is ranked 21st). 
%
Since in practice, the number of incorrect statements is unknown beforehand, developers have to conduct the testing and debugging processes in regression.
After fixing the first bug, developers have to perform regression testing. One more time, \tool effectively ranked the bug in Fig.~\ref{fig:multiple_bugs_example} (\subref{fig:mb_second}) at the $2^{nd}$ position. 
Thus, developers can continuously use \tool in the regression process to quickly find all the bugs in the systems. 
Meanwhile, by using SBFL, two buggy statements are ranked $4^{th}$ and $5^{th}$, respectively.
This shows \textit{the effectiveness of \tool in localizing multiple variability bugs in SPL systems}.

%Explanation or Example
%GPL system
%multiple bugs: _MultipleBugs_.Nob_2.ID_21

\begin{figure}[h]
    % \centering
    
    \begin{subfigure}{0.55\textwidth}
    % \centering
\begin{lstlisting}[language=Java]
public  void preVisitAction(Vertex v){
   if (v.visited != true) {
      v.VertexNumber = --vertexCounter;
      <@\textcolor{ao}{//Patch: v.VertexNumber = vertexCounter++;}@>
   }
}
\end{lstlisting}
         \subcaption{A variability bug in feature \textit{Number}}
        \label{fig:mb_first}
    \end{subfigure}\hfil 
    
    \begin{subfigure}{0.55\textwidth}
    % \centering
\begin{lstlisting}[language=Java]
public void addEdge( Vertex start, Neighbor theNeighbor){
   original( start, theNeighbor );
   if (isDirected != false) {
      <@\textcolor{ao}{//Patch: isDirected == false}@>
      Vertex end = theNeighbor.neighbor;
      end.addWeight( end, theNeighbor.weight );
   }
}
\end{lstlisting}
        \subcaption{Another variability bug in \textit{WeightedWithNeighbors}}
        \label{fig:mb_second}
    \end{subfigure}\hfil 

     \caption{A case of multiple bugs in \textit{GPL}}
      \label{fig:multiple_bugs_example}
\end{figure}{}

%
% The reason of this phenomena is that after fixing one buggy statement, the number of failing products in the tested set is reduced, thus, the isolated suspicious statements set of \tool is narrowed down. Originally, \tool isolated \textbf{503} statements to be suspicious, after one is fixed, for localizing the remaining incorrect statements, this set is reduced to only \textbf{56} statements. Meanwhile, the SBFL technique considers all the executed statements as suspicious, therefore, its search space is upto \textbf{996} statements.
% %
% Secondly, after the first buggy statement is fixed, the remaining failures all caused by the second incorrect statement. Consequently, the incorrect statement is easily distinguished with the correct ones by the failing tests executed by it. 
\subsection{Time complexity (RQ5)}

\label{sec:running_time}
We conducted our experiments on a desktop with Intel Core i5 2.7GHz, 8GB RAM. In 65\% of the cases, \tool took only about 2 minutes to automatically localize buggy statements in each case. On average, \tool spent about 20 minutes on a buggy SPL system. 
%the incorrect statements in a buggy SPL system which is sampled by using \textit{4}-wise coverage for testing. 
Furthermore, we study \tool's running time in different input aspects including the sample size and the complexity of buggy systems in LOCs. For details, see our website~\cite{website}.

Particularly, the running time of \tool gracefully increased when we used more products to localize bugs (no show). This is expected because to localize the variability bugs, \tool needs to examine the configurations of the more sampled products to detect \bpcs. Additionally, \tool analyzes the detected \bpcs in these products to isolate the suspicious statements. Then calculating their suspiciousness scores, as well as ranking these statements. 

% \begin{figure}
%     \centering
%     \includegraphics[width = 0.9\linewidth]{Image/running_time/running_time_loc.pdf}
%     % \caption{The running time of \tool by the systems having different numbers of LOC with a same sample size}
%     \caption{Running time of \tool by LOC}
%     \label{fig:running_time_loc}
% \end{figure}

% However, \textit{\tool's running-time does not linearly increase according to the size of the system}, which is measured by the number of lines of code (LOC). 
% In this experiment, for each case of each system, 8 products were randomly sampled for testing. 

% Fig.~\ref{fig:running_time_loc} shows the average execution time of \tool on the 5 studied systems with different numbers of LOC. 

For the complexity of buggy systems, in general, \tool took more time to analyze the system which has more lines of code  (no show). In particular, \tool took the least time to localize bugs in the smallest system, \textit{BankAccountTP}, while it took more time to investigate a case in the larger one, \textit{ZipMe}. However, \tool needs more time to investigate a case in \textit{Email} compared to a case in \textit{ExamDB}, which is larger than \textit{Email} in terms of LOC. The reason is, \tool analyzes all the failing products to isolate suspicious statements. In our experiment, there are many cases where the number of failing products in \textit{Email} is larger than the number of failing products in \textit{ExamDB}. 

% For detailed results, see our website~\cite{website}.
%OLD
% However, \textit{\tool's running-time does not linearly increase according to the size of the system}, which is measured by the number of lines of code (LOC). In this experiment, for each case of each system, 8 products were randomly sampled for testing. Fig.~\ref{fig:running_time_loc} shows the average execution time of \tool on the 5 studied systems. In general, \tool takes more time to analyze the system which has more lines of code. Particularly, \tool took the least time to inspect a case in the smallest system, i.e., \textit{BankAccountTP}, meanwhile, it took the most time to investigate a case in the largest system, i.e., \textit{GPL}. However, \tool needs more time to investigate a case in the \textit{Email} system compared to a case in the \textit{ExamDB} system, even the \textit{Email} system is smaller. The reason is that \tool needs to analyze all the failing products in the tested set to isolate suspicious statements of the system. In fact, the  tested product sets for the cases of the \textit{Email} system contain more failing products than the tested sets for the cases of the \textit{ExamDB} system. 

\subsection{Threats to Validity}
The main threats to the validity of our work are consisted of
three parts: internal, construct, and external validity threat.

\textbf{Threats to internal validity} mainly lie in the correctness of the implementation of our approach. To reduce this threat, we manually reviewed our code~\cite{website} and verified our program analysis tools' outputs.

\textbf{Threats to construct validity} mainly lie in the rationality of the assessment metrics. To reduce this threat, we chose the metrics that have been recommended by prior studies/surveys~\cite{naish2011model} and widely used in previous work~\cite{keller2017critical, pearson2017evaluating}.

\textbf{Threats to external validity} mainly lie in the benchmark used in our experiments.
The artificial bugs are generated by the mutation testing tool, this will make the diversity of artificial faults in the benchmark, 
%and may not piratically reflect the number of bugs in the systems,
% Particularly, the numbers of faults in the systems in different sizes 
yet could also introduce a bias in our evaluation. 
To reduce this threat, our experiments are conducted on various kinds of mutation operators on both single-bug and multiple-bug settings.

In addition, there is also a threat to external validity is that the obtained results on artificial faults can not be generalized for large-scale SPL systems containing real faults. To mitigate the threat, we chose six widely-used systems in existing studies~\cite{apel2013strategies, interaction_complexity, apel2011language}, which target different application domains, and obtained consistent results on these systems. 
%and they are most  in existing studies for our evaluation, 
%could reduce this threat, these systems still are relatively small scale systems. Consequently, there is a threat to external validity that the obtained results cannot be generalized for other large scale SPL systems. Additionally, although the experimental results of artificial faults have low variance (due to the large number of mutants), they may be biased.
%
Moreover, although it has been very common to evaluate and compare fault localization techniques using artificial faults as a proxy to real faults~\cite{pearson2017evaluating}, it remains an open question whether results on artificial faults are characteristic of results on real faults. 
%
% To mitigate these threats, 
In future work, we are planning to create manually-seeded faults and collect more real-world variability bugs in larger SPL systems to evaluate our technique to address these threats. 
As another external threat, all systems in the benchmark are developed in Java. Therefore, we cannot claim that similar results would have been observed in other programming languages or technologies. This is a common threat of several studies on configurable software systems~\cite{wong2018faster, souto2017balancing}.

Another threat is that our selected SBFL metrics might not be representative. 
% However, we chose a large number of the most popular SBFL metrics. 
To reduce the threat, we chose a large number of the most popular SBFL metrics~\cite{keller2017critical, naish2011model, pearson2017evaluating}.

% Our data has only Java code and not representative. However, we chose multiple projects with large and various numbers of statements and features. Our artificial variability bugs are not the real bugs made by developers. However, our variability bugs are generated based on various kinds of mutation operators.
% Our selected set of SBFL metrics is not representative. However, we chose a large number of the most popular SBFL metrics.

\section{Related Work}
\textbf{Fault Localization}. 
There are various approaches proposed to identify the locations of faults in programs~\cite{wong2016survey, kusumoto2002experimental, tip2001slicing, demillo1996critical}. 
Program slicing~\cite{static_slicing, dynamic_slicing} is used in many studies to reduce the search space while localizing bugs by deleting irrelevant parts in code. Both static slicing~\cite{ kusumoto2002experimental, tip2001slicing} and dynamic slicing~\cite{alves2011fault, demillo1996critical} are used to aid programmers in finding defects.  
%agrawal1993debugging, 
In addition, by SBFL, a program spectrum, which records the execution information of a program, can be used to localize bugs. This idea was suggested by Collofello and Cousins~\cite{collofello1987towards}. To calculate suspiciousness, in early studies~\cite{agrawal1991execution,korel1988pelas}, only information of failed tests was used. 
%taha1989approach
In later studies~\cite{abreu2007accuracy, wong2010family, zhang2017boosting}, which are better, both of the passed and failed tests are utilized. 
Moreover, several studies~\cite{chaleshtari2020smbfl, li2020more} have shown that the FL performance is improved by combining SBFL technique with slicing methods.  
In comparison, while these techniques are designed to localize bugs in a particular product and cannot be directly applied for variability bugs, \tool is specialized for variability bugs in SPL systems.
% Moreover, as we have shown that blindly applying these techniques and not considering the incompatibility of testing in different products could result inaccurate suspiciousness assessment.

\tool is closely related to the work by Arrieta et al. \cite{arrieta2018spectrum}. They use SBFL metrics to localize bugs at the \textit{feature-level} and then create valid products of minimum size containing the most suspicious features. 
% they do not analyze the root causes of the failures of the SPL systems while \tool does
That approach focuses on identifying buggy features where all the containing statements are considered equally suspicious, while \tool is designed to localize variability bugs in the \textit{statement-level}.

% \tool combines and adapts both \textit{spectrum-based} and \textit{slice-based} techniques to localize variability bugs in SPL systems at the \textit{statement level}. 

% First, static slicing is adapted to isolate suspicious statements. It is used to identify interaction implementation of features which suspiciously caused the failures of the systems. Second, SBFL ranking metrics are used to calculate local suspiciousness of statements in each product. Then these local scores are normalized and aggregated accordingly.

\textbf{Feature Interaction}. 
There are a large number of approaches to investigate feature interactions as the influences of the features on the others' behaviors in an unexpected way~\cite{apel2013exploring, garvin2011feature}. Depending on the level of granularity and purposes, various approaches were proposed to detect feature interactions. 
In black-box fashion, Siegmund et al.~\cite{siegmund2012predicting} detects interactions to predict system performance by analyzing the influences of the selected features on the others' non-functional properties. 
Based on the specifications, model checking is also used to check whether the combined features hold the specified properties~\cite{apel2013feature, calder2006feature}. 
In several studies~\cite{ase19prioritization, angerer2015configuration, soares2018varxplorer}, interactions are detected by analyzing code. Nguyen et al.~\cite{ase19prioritization} detect interactions based on their shared program entities. In other studies  \cite{angerer2015configuration, soares2018varxplorer}, they leverage  control and data flow analyses to identify the interactions among features. 
% Soares et al.~\cite{soares2018varxplorer} also leverage the \textit{control and data flow}, however, they compare the execution information of the products to dynamically detect interactions.
%
iGen~\cite{igen} employs an iterative method that runs the system, captures coverage data, processes data to infer interactions, and then creates new products to further refine interactions in the next iteration. 
In \tool, feature interactions are detected for the purpose of identifying statements which suspiciously cause the failures. We detect not only interactions of enabled features but also disabled features, which are the root causes of the failures of the systems.
For a set of enabled features, their interaction is detected by their influences on the other's implementation via control and data dependency. For disabled features, their impacts are estimated by the shared program entities.

\textbf{Variability Bugs}. 
Variability bugs are complex and difficult to be detected by both humans and tools because they involve multiple system features and only be revealed in certain products~\cite{42bugs,98bugs, garvin2011feature,kuhn2004software}.
Abal et al.~\cite{42bugs, 98bugs} analyzed the real variability bugs in several large highly-configurable systems such as Linux kernel and Busybox to understand the complexity and the nature of this kind of bugs. 
In order to make a white-box understanding of interaction faults, Garvin et al.~\cite{garvin2011feature} presented the criteria for interaction faults to be present in systems. Their criteria are about the statements whose (non-)execution is necessary for the failure to be exposed/masked. 
%
% Interestingly, in \cite{98bugs, 42bugs, sampling_comparision, kuhn2004software}, they show that \textit{most of the interaction faults involved by less than 6 features}. 
% Therefore, by testing all 6-ways combination of features is effectively equivalent to exhaustive testing.  
%
% In this work, we concentrate on localizing the variability bugs in the SPL systems at the statement level. To isolate the faults, we identify the sets of features which interact to expose the failures by analyzing the testing results of the products. For efficient, we also consider the interactions of less than 7 features.

\noindent\textbf{Quality Assurance for Configurable Code} 
% Configurable systems create a mechanism to flexibly tailor products to customers' needs. Unfortunately, the large number of features, as well as their mutual interactions make their quality become notoriously difficult to assure. 
There are various studies about variability-aware analysis for the purpose of type checking~\cite{kastner2012type, liebig2013scalable}, %chen2014extending, %
testing~\cite{soares2018exploring, wong2018faster},
%meinicke2016essential
and \textit{control/data flow analysis}\cite{bodden2013spllift, liebig2013scalable}. 
%brabrand2013intraprocedural
In addition, to improve the efficiency of the QA process, several approaches about configuration selection~\cite{greiler2012test, do2014strategies} and \textit{configuration prioritization}~\cite{ase19prioritization,al2014similarity} have been proposed. 
% Several approaches are aimed for testing for configurable
% systems~\cite{cabral2010improving,greiler2012test,cohen2007interaction}.

% For example, Al-Hajjiaji et al.~\cite{al2014similarity} select the next configurations for testing based on the similarity of the configurations with the previously selected one. Nguyen et al.~\cite{ase19prioritization} prioritize configurations based on their number of potential bugs which are measured by analyzing the feature interactions of the system. Moreover, exhaustively testing SPL systems to detect faults is also extremely challenging, therefore, various approaches~\cite{cabral2010improving, cohen2007interaction, greiler2012test, do2014strategies} were also proposed to effectively test these systems. 

% including Variability-aware (VA) analysis, static analysis
% of product lines, sampling, prioritizing, testing for configurable
% systems.

\section{Conclusion}

% Software fault localization is very challenging in SPL systems due to variability bugs which can only be exposed under some combinations of specific feature selections. 
We introduce \tool, a novel approach for localizing variability bugs.
First, to isolate the suspicious statements, \tool analyzes the overall test results and the failing products' code to detect the statements related to the interactions that potentially make the bugs (in)visible in the products. Then, \tool ranks each isolated statement based on two suspiciousness dimensions which are measured by both the overall test results of the products and the detailed results of the test cases which are executed by the statement.
% For a buggy system, \tool isolates and ranks the suspicious statements, which are related to the root causes of the failures of the tested products. 
%
We conducted several experiments on a large dataset of buggy versions of 6 SPL systems in  both single-bug and multiple-bug settings. 
Our results show that in the all 30/30 most popular ranking metrics, \tool's performance in \textit{Rank} is 33\%, 50\% and 95\% better than the state-of-the-art FL techniques, S-SBFL, SBFL, and Arrieta et al.~\cite{arrieta2018spectrum}, for single-bug cases.
Moreover, \tool correctly ranked the buggy statement in +65\% of the cases at the top-3 positions in the resulting lists. 
For the cases of multiple-bug, one-third of the bugs in a buggy system are ranked at top-5 positions by \tool. Especially, in 22\% and 65\% of the cases, \tool is able to effectively localize at least one buggy statement at top-1 and top-5 positions of its ranked lists, respectively. 
From that, developers can iterate the process of bugs detecting, bugs fixing, and regression testing to quickly fix all the bugs and assure the quality of SPL systems.
% \newpage
\section*{Acknowledgments}

% The authors would like to thank the reviewers and associate editor Dr. Michael Pradel, for their detailed comments regarding earlier versions of this paper. 

We would like to thank the anonymous TSE reviewers and editors for their detailed and insightful comments.

%
% The research described has been carried out as part of project AAA.

%
In this work, \textit{Kien-Tuan Ngo} was funded by Vingroup Joint Stock Company and supported by the Domestic Master/ PhD Scholarship Programme of Vingroup Innovation Foundation (VINIF), Vingroup Big Data Institute (VINBIGDATA), code VINIF.2020.ThS.04.

\bibliographystyle{IEEEtran}

\bibliography{13.references}


\newpage
\begin{IEEEbiography}[{\includegraphics[width=1in,clip]{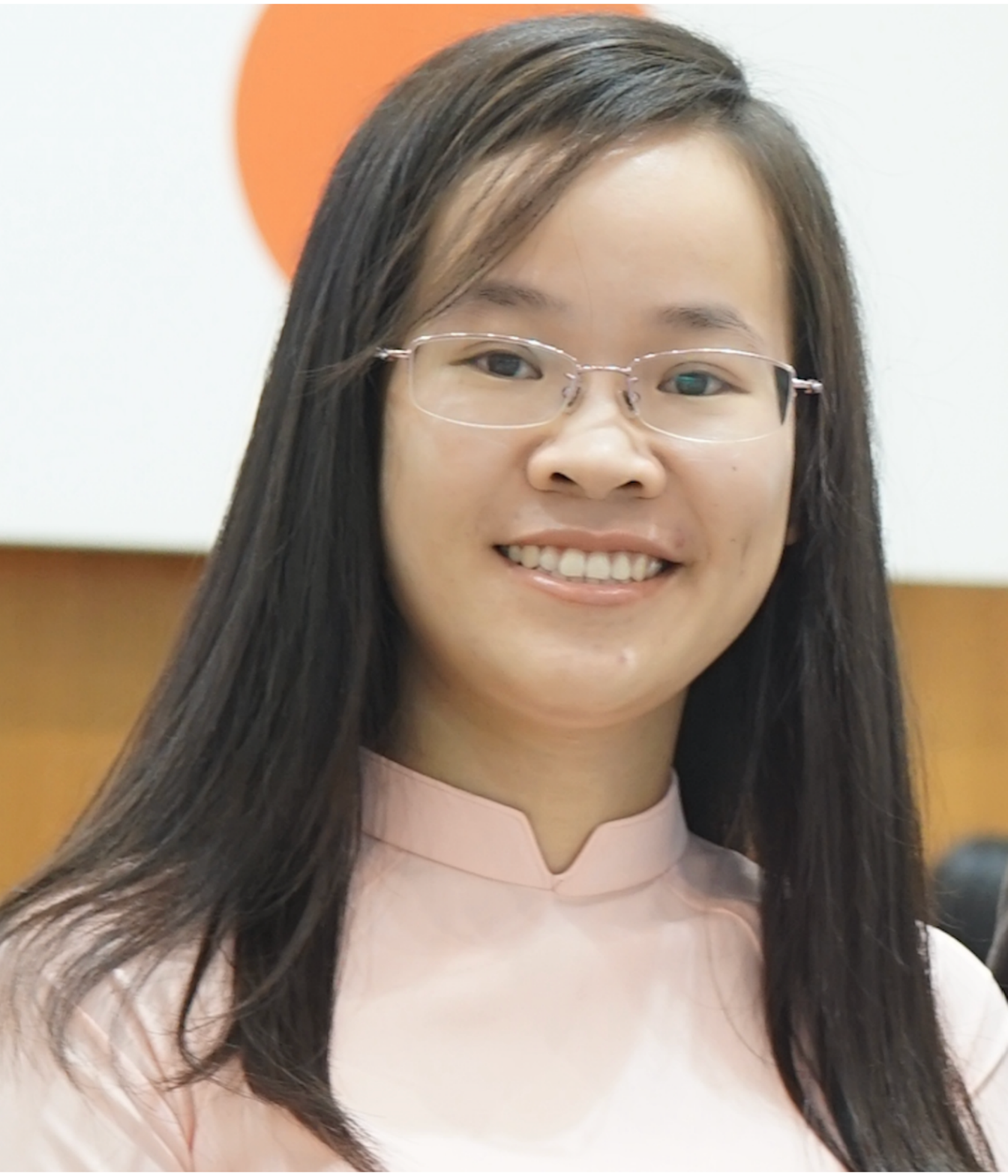}}]{Thu-Trang Nguyen}
is a Ph.D. candidate in Software Engineering at University of Engineering and Technology, Vietnam National University, Hanoi (VNU - UET). She received her MSc degree from Japan Advanced Institute of Science and Technology in 2019 and received her BSc degree from  VNU - UET in 2016.  Her research interests include program analysis, software product line systems, software vulnerabilities detection, and software testing.
\end{IEEEbiography}
\vskip -2\baselineskip plus -1fil
\begin{IEEEbiography}[{\includegraphics[width=1in,clip]{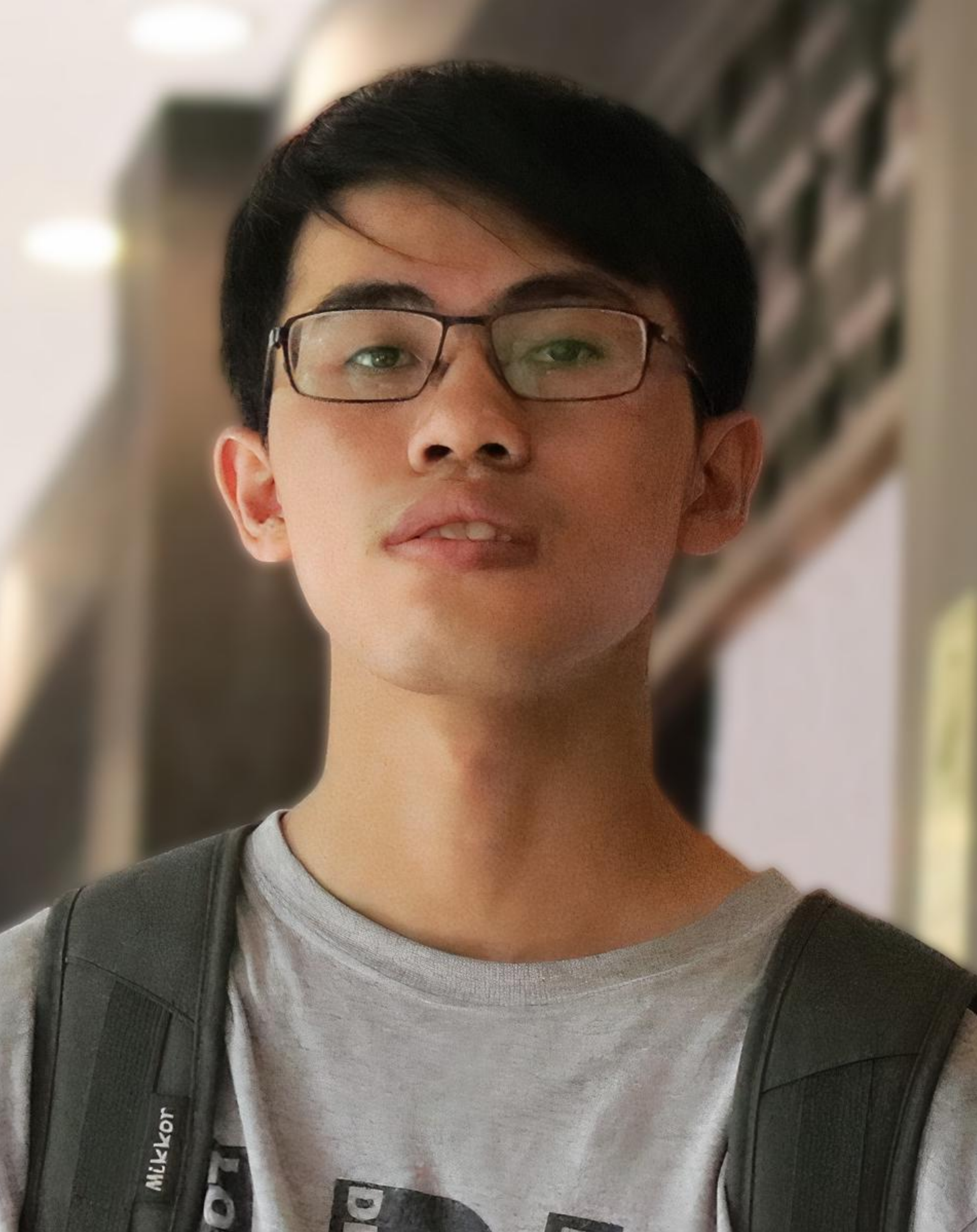}}]{Kien-Tuan Ngo} earned the Bachelor's degree from VNU University of Technology and Engineering, Vietnam in 2020. After that, he continues to work as a full-time researcher at the Faculty of Information Technology at the same college. His research interests include source code analysis, software testing, machine learning for code, and empirical software engineering.
\end{IEEEbiography}

\vskip -2\baselineskip plus -1fil
%%%SON
% \begin{IEEEbiographynophoto}{Son Nguyen}
% is an Ph.D. Candidate in Computer Science at the University of Texas at Dallas.  He received the BSc degree in Computer Science from Vietnam National University, Hanoi, in 2015. His research interests include program analysis, configurable code analysis, and statistical approaches in software engineering.
\begin{IEEEbiography}[{\includegraphics[width=1in,clip]{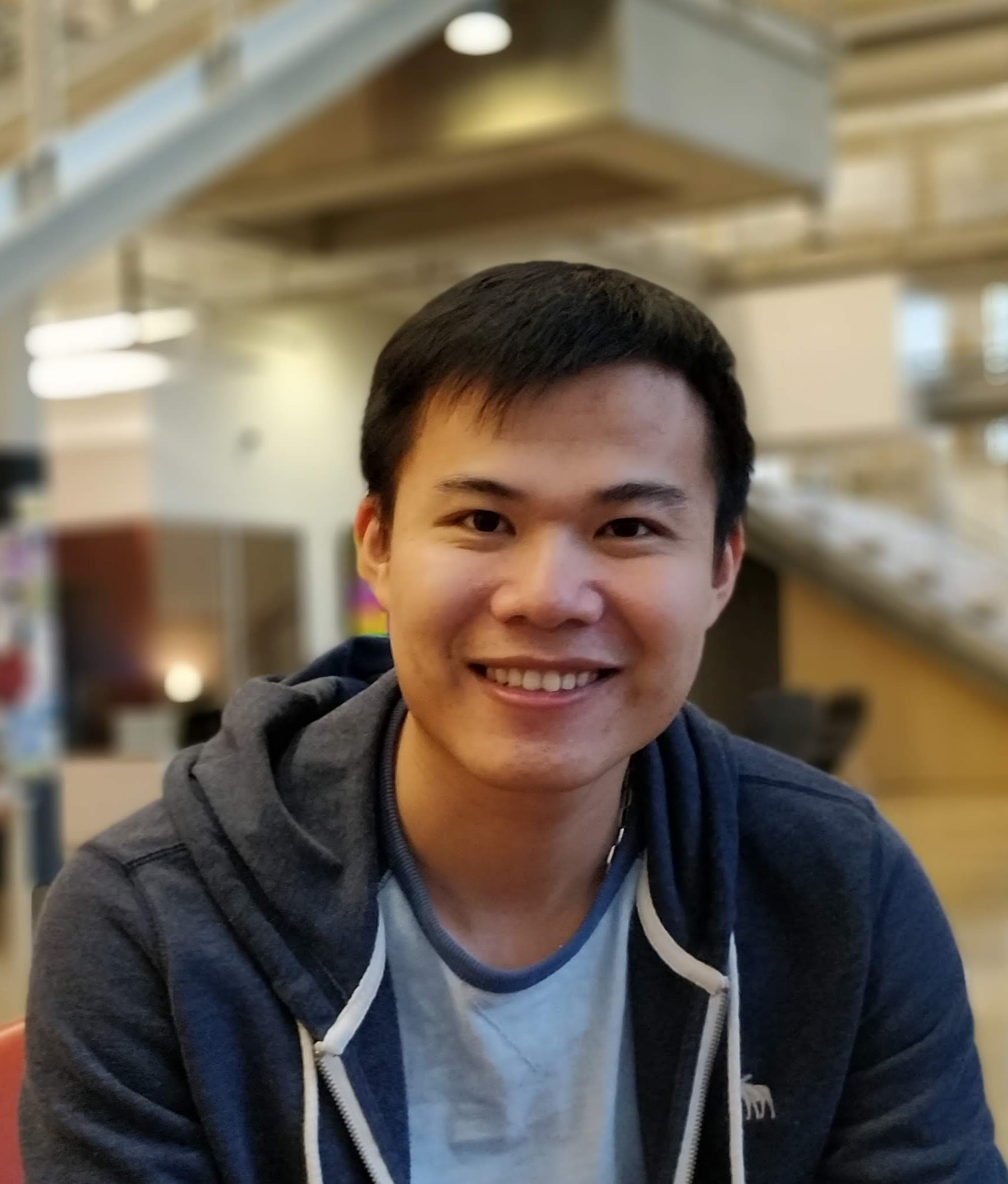}}]{Son Nguyen}
is a Ph.D. Candidate in Computer Science at the University of Texas at Dallas.  He received the BSc degree in Computer Science from Vietnam National University, Hanoi, in 2015. His research interests include program analysis, configurable code analysis, and statistical approaches in software engineering.
\end{IEEEbiography}
%----------------

\vskip -2\baselineskip plus -1fil
\begin{IEEEbiography}[{\includegraphics[width=1in,clip]{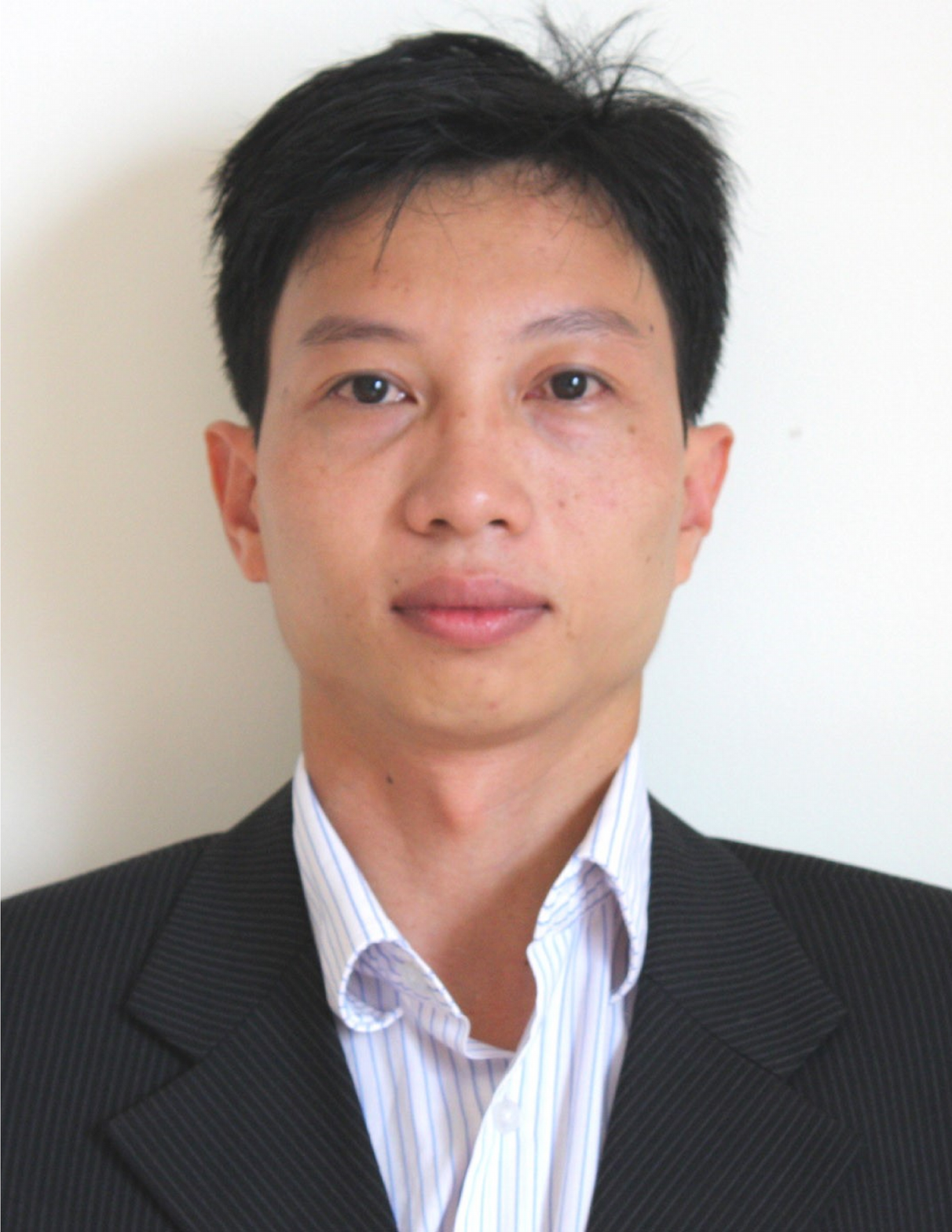}}]{Hieu Dinh Vo}
is the head of the Department of Software Engineering at the Faculty of Information Technology, the University of Engineering and Technology, VNU Hanoi. He earned a Bachelor's degree in Computer Engineering from the Ho Chi Minh City University of Technology, a Master's degree in Distributed Multimedia Systems from the University of Leeds, and a Ph.D. in Information Science from the Japan Advanced Institute of Science and Technology.  His research interests include source code analysis, software testing, service computing, and software architecture.
\end{IEEEbiography}


% % insert where needed to balance the two columns on the last page with
% %\newpage



\vfill

\balance
\end{document}